\pdfoutput=1

\documentclass{sig-alternate-05-2015}
\usepackage{graphicx}
\usepackage{amsmath}
\usepackage{mathpartir}
\usepackage{array}
\usepackage{subfig}
\usepackage{multirow}
\usepackage{multicol}
\usepackage{verbatim}
\usepackage{listings}
\usepackage{color}
\usepackage{hyperref}
\usepackage{url}
\usepackage{dirtree}
\usepackage[nocompress]{cite}
\usepackage{placeins}
\usepackage{enumitem}
\usepackage{balance}
\newcommand{\myapprox}{\raisebox{-0.5ex}{\textasciitilde}}
\newcommand{\R}{R}
\definecolor{cdg}{rgb}{0.0, 0.42, 0.24}

\begin{document}
\pagenumbering{arabic}

\title{ROSA: R Optimizations with Static Analysis}
\numberofauthors{4}
\author{
\end{tabular}
\hspace{-10pt}\begin{tabular}{cccc}
{Rathijit Sen\titlenote{Work done while at UW-Madison}\raisebox{9pt}{$\dagger$}}&
{Jianqiao Zhu\raisebox{9pt}{$\ddagger$}}&
{Jignesh M. Patel\raisebox{9pt}{$\ddagger$}}&
{Somesh Jha\raisebox{9pt}{$\ddagger$}}\\\\
\raisebox{9pt}{$\dagger$}\affaddr{Gray Systems Lab}&
\multicolumn{3}{c}{\raisebox{9pt}{$\ddagger$}\affaddr{Department of Computer Sciences}}\\
\affaddr{Microsoft Corporation}&\multicolumn{3}{c}{\affaddr{University of Wisconsin-Madison}}\\
\texttt{rathijit.sen@microsoft.com}&\multicolumn{3}{c}{\texttt{\{jianqiao,jignesh,jha\}@cs.wisc.edu}}
}

\maketitle


\begin{abstract}
{\R} is a popular language and programming environment for data scientists. It is increasingly co-packaged with both relational and Hadoop-based data platforms and can often be the most dominant computational component in data analytics pipelines. Recent work has highlighted inefficiencies in executing {\R} programs, both in terms of execution time and memory requirements, which in practice limit the size of data that can be analyzed by {\R}. This paper presents ROSA, a static analysis framework to improve the performance and space efficiency of R programs. ROSA analyzes input programs to determine program properties such as reaching definitions, live variables, aliased variables, and types of variables. These inferred properties enable program transformations such as C++ code translation, strength reduction, vectorization, code motion, in addition to interpretive optimizations such as avoiding redundant object copies and performing in-place evaluations. An empirical evaluation shows substantial reductions by ROSA in execution time and memory consumption over both CRAN R and Microsoft R Open.

\end{abstract}


\section{Introduction}
\label{sec:intro}

{\R} is a popular programming language for data analysis~\cite{ihaka:r-language:jcgs:1996,morandat:r-lang-design:ecoop:2012,smith:r-ecosystem:useR:2011,venkataraman:presto-sparse-matrices:eurosys:2013}.
It is the most popular data mining tool~\cite{rexer:data-miner-survey:2013}, and is the third-most used data analysis language after SQL and Excel~\cite{king:data-science-survey:oreilly:2015}. {\R} is also nearly always co-packaged/embedded with Hadoop and relational data processing platforms (e.g.,~\cite{rhadoop, oracle-r-enterprise,ibmdmR,pivotalr,SAP-R,Tibco-R,revolution-analytics-teradata}), making it a crucial part of contemporary data analytics workflows. Given the close integration of {\R} and databases, speeding up {\R} has been a recurring topic in the database research community~\cite{sridharan:profiling-contemporary-processor:vldb:2014,zhang:riot-i/o-efficient:cidr:2009,zhang:riot-io-efficient:icde:2010}, and this research follows that line of thinking.

{\R} has a dynamic, lazy, functional, object-oriented language semantics~\cite{morandat:r-lang-design:ecoop:2012}, and is interpreted~\cite{r:cran}. Although highly expressive, interpretive execution of {\R} programs has space and runtime inefficiencies~\cite{sridharan:profiling-contemporary-processor:vldb:2014,kotthaus:analyses-machine-learning:jscs:2015} that are overwhelming when analyzing large datasets~\cite{sridharan:profiling-contemporary-processor:vldb:2014}, limiting the size of the datasets that can be analyzed with R.

For example, consider the Simple Arithmetic program in Listing~\ref{lst:sa} that computes the distances from a given point to a list of points. This program uses two lists (\texttt{x} and \texttt{y}) of 1 Billion elements (\texttt{n <- 1e9}) each. Increasing the list sizes to 9 Billion elements causes the {\R} interpreter to abort evaluation on our system with 256 GB of memory, as the program runs out of memory. This behavior is surprising since the two lists have a total of 18 Billion (8 byte) double-precision elements, thus requiring less than 140 GB of main memory. However, when we run this program on a machine with 256 GB of main memory, the program crashes as it runs out of memory space. There are significant overheads in the {\R} interpreter, and these issues surface prominently when {\R} code is packaged with data platforms that manage large datasets. Thus, improving the behavior and performance of R programs is crucial for contemporary data platforms.

\begin{lstlisting}[frame=single,frameround=tttt,caption={Simple Arithmetic in R (program from~\protect\cite{zhang:riot-i/o-efficient:cidr:2009,sridharan:profiling-contemporary-processor:vldb:2014}}),label=lst:sa]
n <- 1e9
xs <- 0.5
ys <- 0.5
x <- runif(n)
y <- runif(n)
d <- sqrt((x-xs)^2+(y-ys)^2)
\end{lstlisting}

The focus of this paper is on exploring if the limitations of R discussed above can be mitigated by using compiler techniques (such as~\cite{matloff:art-programming:book:2011,templelang:advanced-compilation-tools:ss:2014,wickham:advanced-r:book:2014}). To the best of our knowledge there hasn't been any previous study that catalogs the list of potentially applicable compiler techniques that are applicable in this setting, and systematically determines which of these can be made to work synergistically with each other and with the idiosyncrasies that come with the R language. 
A key contribution of our paper is addressing this gap. Table~\ref{table:gains} shows a list of static analyses and corresponding optimizations that we have developed in this paper to address this research question. 

\begin{table*}[ht]
\def\arraystretch{1.2}
\centering
\caption{Static analyses techniques, with the associated Optimizations in square brackets, for our workloads. For example, the Simple Arithmetic program is improved by the Space Reuse optimization which is enabled by live variable and alias analyses.}
\begin{tabular}{|c|c|c@{ }|}
\hline
\textbf{Analysis [Optimizations]}& \textbf{Workload}& \textbf{Status}\\\hline
\multirow{7}{*}{Type Inference [Translation to C++ code and compilation]}& Binary Search & {Automated}\\\cline{2-3}
& 2D Random Walk & {Automated}\\\cline{2-3}
& Euclidean Distance & {Automated}\\\cline{2-3}
& OddCount & {Automated}\\\cline{2-3}
& Exponential Smoothing & {Automated}\\\cline{2-3}
& Discrete Value Time Series, ver. A & {Automated}\\\cline{2-3}
& Discrete Value Time Series, ver. B & {Automated}\\\hline
Live Variable and Alias Analyses [Space Reuse] & Simple Arithmetic &{Automated}\\\hline
Reaching Definitions Analysis [Vectorization + Code Motion]& Simple Vectorization &{Automated}\\\hline
Type Inference [Strength Reduction (Float $\rightarrow$ Int)] & \multirow{3}{*}{Unique Genotypes Test} &User-Input\\\cline{1-1}\cline{3-3}
Type Inference [Strength Reduction (Float $\rightarrow$ String)] &  &User-Input\\\cline{1-1}\cline{3-3}
Type Inference [Strength Reduction (Float $\rightarrow$ String) + Code Motion] &  &User-Input\\\hline
Type Inference [Strength Reduction (DataFrame $\rightarrow$ Matrix)] & Kmeans  &User-Input\\\hline
Loop Analysis [Loop Tiling]& Matrix Multiplication &User-Input\\\hline
\end{tabular}
\label{table:gains}
\end{table*}

Implementing these techniques can be challenging. One can certainly build each technique individually as a standalone technique/package, but a better way is to incorporate these techniques as first-class analyses into a compiler. This integrated compiler-based approach is what we take in this paper, creating an R-optimization framework called ROSA. This integrated approach enhances ease-of-use for the end user, and also enables reuse of analysis results across optimizations; e.g., type inferencing results can be used for vectorization, strength reduction, and code translation. 

To illustrate, with our Space Reuse optimization, the {\R} interpreter can process the Simple Arithmetic program discussed above with 18 billion elements.

\begin{figure}[bt]
\centering
\includegraphics[width=0.35\textwidth]{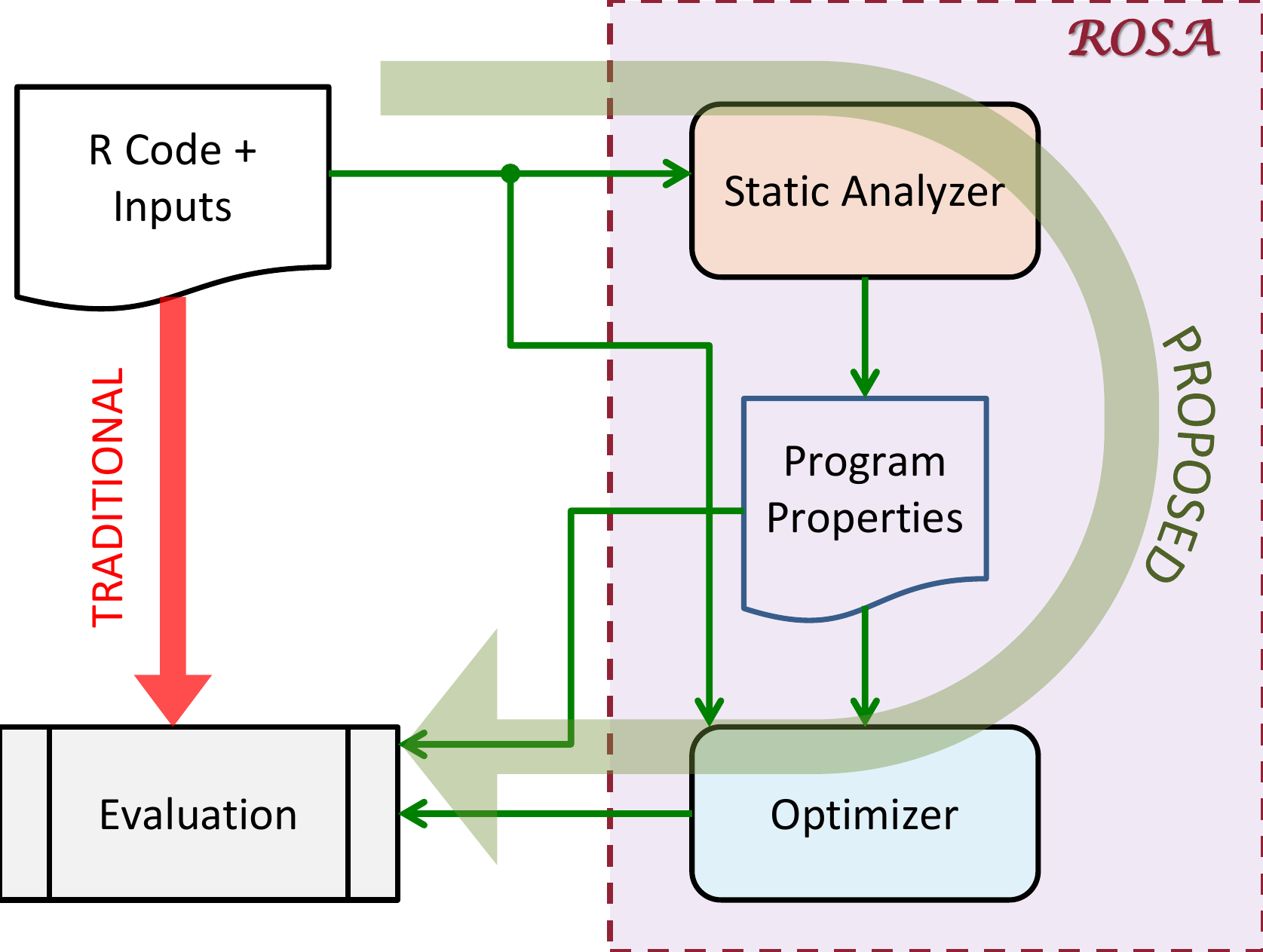}
\caption{System Architecture}
\label{fig:architecture}\vspace{-10pt}
\end{figure}
A crucial aspect of our design is that large parts of it can work without requiring modifications to the existing R programs. This aspect is important as a big driver behind the popularity of R is the large code base of user code and CRAN packages that are already deployed. Our approach is to use a compiler-based approach as outlined in Figure~\ref{fig:architecture}.

The left side of Figure~\ref{fig:architecture} shows a high-level schematic of the traditional workflow for executing {\R} programs. The right side of this figure shows the new workflow with ROSA. In the traditional workflow, the {\R} interpreter directly evaluates the given {\R} program and its inputs. During evaluation it executes any precompiled portions directly on the host machine. In ROSA, the program along with its inputs is first analyzed by the \textit{Static Analyzer} to determine various properties at each point in the program. These properties can then be inspected by the ROSA \textit{Optimizer} for various transformations, e.g., vectorization, code motion, C++ code translation, etc. This optimized code is then evaluated. The inferred program properties are also used during evaluation by the {\R} interpreter to avoid making redundant object copies and perform in-place computations, if possible.

Our system automates the optimizations shown in Table~\ref{table:gains} that are marked with the status tag `Automated'; i.e. these techniques work without requiring any changes to existing R programs. Due to inherent characteristics of the R language (discussed further in Section~\ref{subsec:experiments-improvements-baseline}), the remaining optimizations require user feedback. However, we note that even unmodified program can benefit dramatically with ROSA, as demonstrated by the results presented in Section~\ref{sec:experiments}.

The key contributions of this paper are as follows.

\begin{enumerate}
\item We propose the ROSA framework that integrates compiler-based optimization techniques with {\R}'s evaluation framework. We demonstrate using an empirical evaluation that ROSA improves performance, and has a smaller memory footprint compared with both Microsoft R Open and CRAN R. 
\item We propose a type inferencing system that generates crucial information necessary for automated translation of {\R} programs to efficient C++ code. While speeding up of {\R} programs using C++ code is known, type inferencing for \textit{automated} translation into C++ code has not been hitherto explored. Type inferencing information is also useful for vectorization and strength reduction transformations.
\item We show how enhancements to the {\R} interpreter, that utilize live variable analysis and alias analysis information, can overcome the space inefficiencies of the existing copy-on-write policy of {\R}. We also highlight the importance of strength reduction transformations in improving performance of {\R} programs by reducing or eliminating costly type-conversion operations.
\end{enumerate}

The remainder of this paper is organized as follows. Section~\ref{sec:r-inefficiencies} presents the required preliminaries. Section~\ref{sec:rosa-architecture} presents a more detailed architecture of ROSA.
Section~\ref{sec:sa} describes relevant static analysis techniques. Section~\ref{sec:optims} describes the optimizations that use the inferred program properties. Section~\ref{sec:experiments} presents empirical results. Section~\ref{sec:related-work} covers related work and Section~\ref{sec:conclude} contains our concluding remarks.
The Appendix includes code for the {\R} programs that we use.


\section{Background and Key Issues}
\label{sec:r-inefficiencies}

Next, we present background related to how {\R} works. In addition, to motivate the optimizations discussed in Section~\ref{sec:optims}, we also 
discuss the key causes for space and time 
inefficiencies in the {\R} interpreter.

\subsection{Interpretation}
\label{subsec:ineff-interpret}

{\R} is an interpreted language. Program statements are interpreted by an \texttt{eval} function that looks up values of symbols and implementation of operators while evaluating the expression in the given environment. Interpretation of every operator results in a call to a function that implements the operator. For example, an expression of the form \texttt{a+b} will result in a lookup for the \texttt{+} operator and the internal function \texttt{do\_arith} will be called to perform the operation. Expressions such as \texttt{a[x]} result in calling the \texttt{do\_subset} function or the \texttt{do\_subassign} function. Repeated lookups and function calls can be very expensive, particularly when they happen repeatedly, e.g, in loops. This interpretive nature of R results in the creation of a large number of temporary variables that can have a very high execution overhead. An example illustrating this issue is shown in Appendix~\ref{sec:detailed-overhead-interpretation}. 

Compiling {\R} code to C/C++ is a known technique to improve efficiency~\cite{eddelbuettel:rcpp:JSSOBK:2011,templelang:advanced-compilation-tools:ss:2014,garvin:rcc-compiler:rice:2004,eddelbuettel:rcpparmadillo:csda:2014,talbot:riposte-compiler-parallel:pact:2012}. Generating C/C++ code and executing compiled code leads to significant performance improvements to {\R} programs. One important challenge in \textit{automatic translation} of {\R} programs to C/C++ is to statically determine variable types in the program/target subroutine so that proper declarations, object iterators, and access methods can be generated. Translation to C/C++ is not possible without type information. Simple type inferencing can be helpful, as illustrated in Appendix~\ref{sec:detailed-overhead-interpretation}.

\subsection{Copy-on-write Semantics} 
\label{subsec:ineff-cow}

In an assignment of the form \texttt{y <- x}, both \texttt{x} and \texttt{y} point to the same memory location, unless one of them is written to, in which case a copy is made. However, a copy may not be needed if the other variable is not live beyond that point; i.e., its value is no longer needed. Restricting copies can save memory space especially when dealing with large objects, such as long vectors.

\vspace*{-1ex}
\begin{lstlisting}[frame=single,frameround=tttt,caption={Copy-On-Write},label=lst:cow]
n <- 1e8
x <- rep(1,n)
y <- x
x[2] <- 3
y[2] <- 3
\end{lstlisting}

The Copy-On-Write example, Listing~\ref{lst:cow}, illustrates how live variable analysis can reduce memory overheads. Due to the copy-on-write semantics of {\R}, the assignment on line 3 does not create a new allocation, but the assignments on lines 4 and 5 do. The {\R} interpreter internally maintains a \textit{named} field for every S-expression. The value of this field can be 0 (not shared), 1 (internal use), or 2 (may be shared). The assignment on line 3 sets the named field to 2 for the object pointed to, in this case, the long vector. The assignments on lines 4 and 5 notice that there may be a shared value, and creates copies of the vector with the \textit{named} field set to 0 in the copies. The copy on line 5 can always be avoided, but {\R} does not track the set of variables that point to the same object. Hence, it cannot determine that after line 4, \texttt{x} and \texttt{y} are no longer aliased. Moreover, if \texttt{x} and \texttt{y} no longer live beyond line 4, then the copy on line 4 can be avoided. In fact, the assignment on line 4 need not be performed in this example. 

Instead of copy-on-write, a more efficient semantic would be to have copy-on-write\emph{-and-live-sharers}. That is, a copy is needed during modification of an aliased object only if some of the other aliases may live beyond that point.

\subsection{Attribute Evaluations}
\label{subsec:ineff-attrib}

{\R} maintains attributes (meta-data) for each object. Some important attributes are ``class'' (class of the object) used by a dispatch function, ``dim'' (dimension) used for arrays and matrices, ``dimnames'' (names of dimensions), ``rownames'', ``colnames'', ``names'', and ``tsp'' used for time-series objects.

\begin{lstlisting}[frame=single,frameround=tttt,caption={Kmeans~\protect\cite{sridharan:profiling-contemporary-processor:vldb:2014}},label=lst:kmeans]
A <-read.table(file="airline150M.csv", sep=",", header=T, nrows=149545445,...)
gc(T)
system.time(result <- kmeans(na.omit(A),2,iter.max=1000,algorithm="Lloyd"))
gc(T)
\end{lstlisting}

Some computation is performed by the {\R} interpreter to maintain attributes during interpretation of an {\R} program. Depending on the size of the object and the attribute, this step can be quite costly. An illustrative example of this overhead for the Kmeans program, Listing~\ref{lst:kmeans}, is shown in Appendix~\ref{sec:detailed-overhead-attribute} where we discuss how implicit conversion from a dataframe object to a matrix can be inefficient. In Section~\ref{sec:experiments} we present a reduction transformation to avoid this overhead.

\begin{lstlisting}[frame=single,frameround=tttt,caption={Unique Genotypes Test~\protect\cite{reynolds:genetic-equilibrium-conservation:phdthesis:2011,reynolds:ugt-r-script:2011}},label=lst:ugt]
NG.test <- function(X,N,n,reps){
  L <- length(X) 
  G <- numeric()
  for(i in 1:reps){
    genos <- matrix(NA,N,L)
    for(j in 1:L){
    genos[,j] <- sample(c(0,1),size=N,replace=TRUE,prob=c(1-X[j],X[j]))
    }
    geno.c <- numeric()
    for(j in 1:N){
      geno.c[j] <- paste(genos[j,],sep="",collapse="")
    }
    G[i] <- length(unique(geno.c))
    }
  G
}

X <- rbeta(29,.2,.2)
N <- 29
n <- 15
reps <- 100000
system.time(xx <- NG.test(X=X,N=N,n=n,reps=reps))
\end{lstlisting}

\subsection{Type Conversions (particularly, ToString)} 
\label{subsec:ineff-conversion}

{\R} statements consist of one or more S-expressions that can be of different types such as integer, string, list, etc. A lot of time can be lost due to conversion between types. 

The Unique Genotypes Test~\cite{reynolds:ugt-r-script:2011,reynolds:genetic-equilibrium-conservation:phdthesis:2011}, Listing~\ref{lst:ugt}, samples values from 0 and 1, and then finds the number of unique patterns. A key time-consuming operation in this program is the \texttt{paste} operation, internally implemented by {\R} using the \texttt{do\_paste} function. This function converts its arguments into strings, if they are not already strings. The elements 0 and 1 passed to the \texttt{sample} function on line 7 in Listing~\ref{lst:ugt} are floats. Conversion from float to string is expensive. Changing the inputs to ``0'' and ``1'' causes them to be treated as strings and avoids the type conversion altogether. The Kmeans example discussed in Section~\ref{subsec:ineff-attrib} also suffers from type conversion overheads caused by attribute evaluations. 

Knowledge of how the inputs will be used including type information can help to identify strength reduction opportunities that can reduce/eliminate these conversion overheads.

\subsection{Memory Management} 
\label{subsec:ineff-memory}

During the course of evaluation of S-expressions, memory for temporary and program variables are allocated by the memory allocator and reclaimed by the garbage collector when not needed. Memory management can be expensive if thresholds are not properly set~\cite{sridharan:profiling-contemporary-processor:vldb:2014}.

The Simple Arithmetic program, Listing~\ref{lst:sa}, highlights an overhead discussed in prior work~\cite{zhang:riot-i/o-efficient:cidr:2009,sridharan:profiling-contemporary-processor:vldb:2014}. The issue here is that a new allocation is made for the intermediate result $\mathtt{x-x_s}$ and another one for $\mathtt{y-y_s}$. This allocation step can use up a lot of space if \texttt{x} and \texttt{y} are large vectors. However, if we can utilize the information that \texttt{x} and \texttt{y} are not needed after this computation, then their memory spaces can be reused for computation. This determination can be enabled by \textit{live-variable analysis}---neither \texttt{x} nor \texttt{y} are live beyond this point.

The following example shows another scenario where allocations can be avoided with in-place computations if variables are no longer live beyond that point. The statement \texttt{x[-1]} returns an object with all except the first element of \texttt{x}, and can reuse the space of \texttt{x} if the original object \texttt{x} is not needed again.

\begin{lstlisting}[frame=single,frameround=tttt]
n <- 1e8
x <- as.double(sample(1:100,n,TRUE)
y <- x[-1]
\end{lstlisting}

Avoiding allocations for large temporary objects helps to reduce the maximum memory resident set size (RSS). This is important since if RSS exceeds the available physical memory on the system, thrashing will happen leading to higher (swap) disk I/O and significantly reduced performance. A larger RSS also reduces the memory that is available to other applications, such as a co-packaged database server/service.

Just having live information available during interpretation is not sufficient. The {\R} interpreter should also have access to operation implementations that leverage liveness information to compute in-place and reduce memory usage.

\begin{figure*}[ht]
\centering
\includegraphics[width=0.95\textwidth]{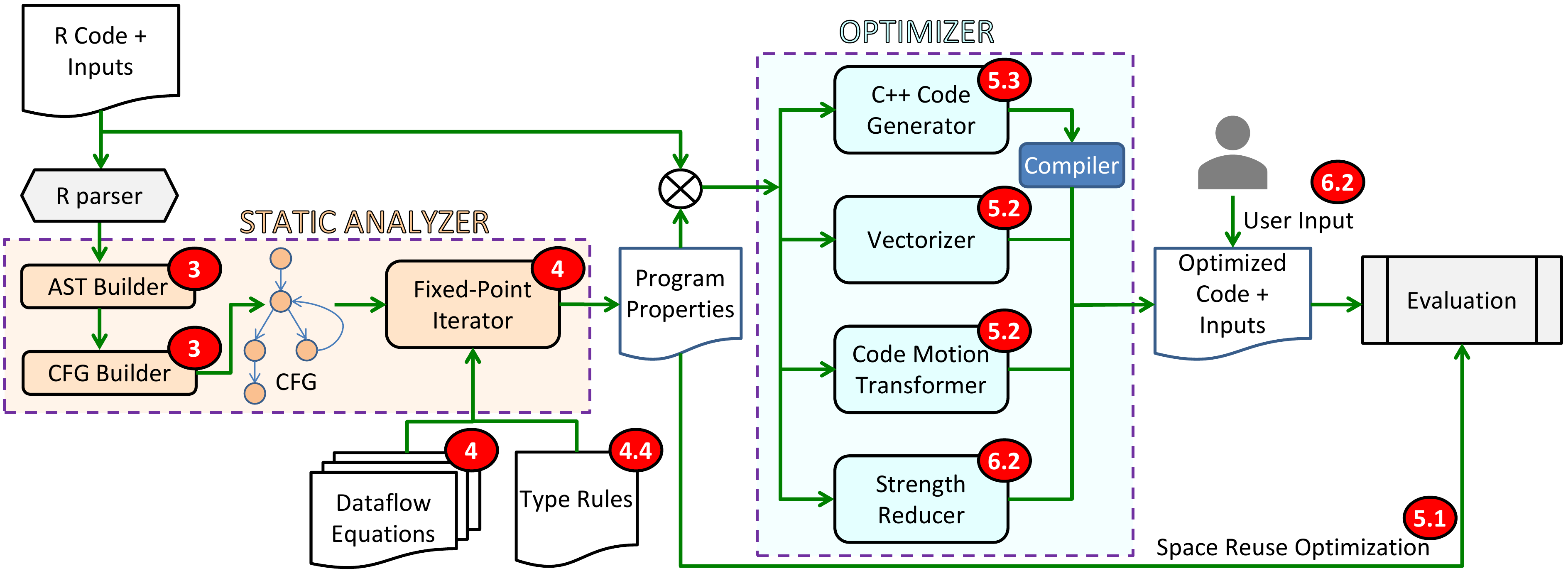}
\caption{ROSA architecture. Components are tagged by shaded ovals with the section number where that aspect is discussed.}
\label{fig:architecture-detailed}
\end{figure*}

\subsection{Non-Vectorized Computations}
\label{subsec:ineff-repeated}

Repeated computations during interpretation of statements in loops cause overheads. Vectorization is a well-known technique to improve {\R} performance~\cite{matloff:art-programming:book:2011}. For example, the Simple Vectorization program, Listing~\ref{lst:vectorization}, adds two vectors element by element in a loop (\texttt{x[i]=y[i]+z[i]}). The repeated interpretation within the loop can be avoided by removing the loop and using vectorized addition (\texttt{x=y+z}), leading to significant performance benefits. Automatically detecting opportunities for vectorization requires determining \textit{loop induction variables} and absence of \textit{loop-carried dependencies}.

\begin{lstlisting}[frame=single,frameround=tttt,caption={Simple Vectorization~\protect\cite{matloff:art-programming:book:2011}},label=lst:vectorization]
n <- 1e8
x <- runif(n)
y <- runif(n)
z <- vector(length=n)
system.time(for(i in 1:n) z[i] <- x[i] + y[i])
\end{lstlisting}

\begin{lstlisting}[frame=single,frameround=tttt,caption={2D Random Walk~\protect\cite{templelang:advanced-compilation-tools:ss:2014}},label=lst:2dwalk]
rw2d1 = function(n = 100) {
  xpos = numeric(n)
  ypos = numeric(n)
  for(i in 2:n) {
    delta = if(runif(1) > .5) 1 else -1
    if (runif(1) > .5) {
      xpos[i] = xpos[i-1] + delta
      ypos[i] = ypos[i-1]
    }
    else {
      xpos[i] = xpos[i-1]
      ypos[i] = ypos[i-1] + delta
    }
  }
  return(list(x = xpos, y = ypos))
}

n = 1e7
system.time(b <- rw2d1(n))
\end{lstlisting}

The 2D Random Walk example, Listing~\ref{lst:2dwalk}, incurs overhead due to repeated calls to the random number generator function, \texttt{runif}. On every call, an internal function requests the memory allocator to allocate space to copy the working state of 624 integers for the Mersenne Twister pseudo-random number generator~\cite{matsumoto:mersenne-twister-prng:tomacs:1998}. However, \texttt{runif} can be vectorized to generate multiple random numbers in a single call (\texttt{runif(n)}), resulting in a single allocation request, and hoisted out of the loop. To enable this vectorization, loop index variable and loop bounds analyses are needed to determine the arguments to pass to a vectorized \texttt{runif}. 

\subsection{Cache Access Patterns}
\label{subsec:ineff-cache}

\begin{lstlisting}[frame=single,frameround=tttt,caption={Matrix Multiplication~\protect\cite{sridharan:profiling-contemporary-processor:vldb:2014}},label=lst:matmul]
n <- 1073741824
A <- matrix(sample(c(1:100,NA),n,T),ncol=4194304)
B <- matrix(sample(c(1:100,NA),n,T),nrow=4194304)
gc(T)
system.time(C <- A %*% B)
gc(T)
\end{lstlisting}

{\R} interprets matrix data in column-major order. However, implementations of some operations, e.g., matrix multiplications of matrices with \texttt{NA} values (as in the Matrix Multiplication program, Listing~\ref{lst:matmul}) cause memory accesses with poor cache locality. The {\R} implementation of the matrix multiplication operator \texttt{\%*\%} accesses matrix \texttt{A} in row-order and matrix \texttt{B} in column-order. A more cache-friendly access pattern is to access \texttt{A} in column-order as well. Changing the access order can reduce overheads due to cache misses. Analyses that determine data dependencies and reuse across loop iterations~\cite{wolf:data-locality-optimization:pldi:1991} can help identify these opportunities.


\section{ROSA Architecture}
\label{sec:rosa-architecture}

Figure~\ref{fig:architecture-detailed} shows a detailed view of the architecture of ROSA. The {\R} program, along with its inputs, are first parsed by the {\R} parser (available as part of the standard {\R} package) to create the internal representation (IR) of the code. This IR is a list of structures called S-expressions that represent the syntax tree of the given program. This is the same IR that the {\R} interpreter uses during traditional evaluation. 
This IR is then processed by the \textit{Static Analyzer} as follows:
\begin{enumerate}
\item The \textit{AST Builder} applies transformations to the original syntax tree (e.g., transforms \texttt{if}/\texttt{for}/\texttt{while} statements and function definitions into ``normal" forms, for ease of later processing), annotating it with program structure information (e.g., loop bodies).
\item The \textit{CFG Builder} recursively processes the syntax tree to extract the Control Flow Graph (CFG) for the program with each statement as a node in the graph.
\item The \textit{Fixed-point Iterator} applies data flow equations and type inferencing rules to determine program properties that can be used to decide how to optimize the given program. Section~\ref{sec:sa} discusses these analyses in detail.
\end{enumerate}
Appendix~\ref{sec:example-analysis} shows a detailed example of the IR, CFG, and inferred types for a program.

Next, the \textit{Optimizer} uses the inferred properties to apply various transformations as follows:
\begin{itemize}
\item The \textit{C++ Code Generator} uses the results of type inferencing to create (efficient) C++ code. This code is then compiled using g++ to create a shared library. This library will be invoked during evaluation using {\R}'s ``.Call'' interface. Currently we generate code only for a subset of {\R} types.
\item The \textit{Vectorizer} identifies vectorization opportunities, e.g., in the Simple Vectorization code, Listing~\ref{lst:vectorization}.
\item The \textit{Code Motion Transformer} identifies loop-invariant computations and hoists them outside of the loop. This can happen, e.g., due to vectorization (vectorized code is moved out of the loop), repeated allocations (such as in the Unique Genotypes Test, Listing~\ref{lst:ugt}), etc.
\item The \textit{Strength Reducer} identifies opportunities for substituting data types in the input code to other data types for more efficient processing, e.g., in the \texttt{Kmeans} program, Listing~\ref{lst:kmeans}, and Unique Genotypes Test.
\end{itemize}

Section~\ref{sec:optims} discusses these optimizations in more detail. Currently, a few of these optimizations, e.g., strength reduction, require user input for their transformations. 
Table~\ref{table:gains} shows the current status of the optimizations.

Apart from the transformations, the program properties determined by the static analyzer is used by the {\R} interpreter to avoid unnecessary object copies so that existing allocated space can be reused, e.g., for the Simple Arithmetic program, Listing~\ref{lst:sa}. Section~\ref{subsec:optims-space-reuse} discusses this optimization.


\section{Static Analyses}
\label{sec:sa}

Once the CFG is available, the static analyzer iterates over the CFG for each analysis and determines facts that are true at the entry and exit nodes of each basic block. The iterations terminate when a fix-point is reached; i.e., facts do not change on further application of analysis rules. The same facts are generated regardless of the order of basic blocks considered in every iteration. 
We will now briefly describe the key analyses.

\subsection{Live Variables Analysis}
\label{subsec:sa-live-var-analysis}

The goal of this analysis is to determine the set of variables that are live at each program point. Variables that are not live (i.e., dead) do not have their values used later in the program before possible re-definition. Space allocated for dead variables can be used for other purposes. This analysis can help the {\R} interpreter avoid (or reduce) making unnecessary memory allocations for objects.

Let $live\_in(j)$ denote the set of variables that are live; i.e., their values will be used before re-definition, at the entry of basic block $j$. $live\_in(j)$ is augmented by $gen(j)$ which is the set of generated variables; i.e., the set of variables appearing in the RHS of the statements in the basic block. $gen(j)$ is thus the set of variables whose values are used before re-definition in this basic block. $live\_in(j)$ is reduced by $kill(j)$ which is the set of killed variables; i.e., the set of variables appearing in the LHS of the statements in the basic block. $kill(j)$ is thus the set of variables whose values are re-defined in this basic block.

Let $live\_out(j)$ denote the set of variables that are live at the exit of the basic block $j$. This is equal to the union of all variables that are in $live\_in$ at the entry points of successor basic blocks.

The analysis equations are given below. This is a backward dataflow analysis.
\begin{eqnarray*}
live\_in(j) &=& gen(j)\bigcup\,(live\_out(j)-kill(j))\\
live\_out(j) &=& \bigcup_{i \in succ(j)} live\_in(i)
\end{eqnarray*}

\subsection{Alias Analysis}
\label{subsubsec:sa-alias-analysis}

At a given point in a program, a variable can point to one of a number of possible locations. An alias set is a set of variables with at least one mutually common location that they may point to. The goal of this analysis is to determine alias sets. We perform a flow-sensitive analysis~\cite{hardekopf:ptranalysis-millions-loc:cgo:2011,woo:flow-sensitive-alias:hpc:2000}. 

Let $alias\_out(j)$ denote the set of alias sets at the exit of basic block $j$. $alias\_out(j)$ is augmented by $gen(j)$ which is the set of aliases generated by copy assignments such as \texttt{x=y}, \texttt{x<-y}, or \texttt{x{<}<-y} in the basic block. $alias\_out(j)$ is reduced by $kill(j)$ which is the set of killed variables, that is, the set of variables appearing in the LHS of statements that are not copy assignments in the basic block. $kill(j)$ is thus the set of variables that no longer have earlier alias relations after this basic block. Every variable in $kill(j)$ is removed from all alias sets in $alias\_in(j)$.

Let $alias\_in(j)$ denote the set of aliases at the entry of the basic block $j$. This set is equal to the union of all alias sets that are in $alias\_out$ at the exits of the predecessor basic blocks.

The analysis equations are given below. This is a forward dataflow analysis.
\begin{eqnarray*}
alias\_in(j) &=& \bigcup_{i \in pred(j)} alias\_out(i)\\
alias\_out(j) &=& gen(j)\bigcup\,(alias\_in(j)-kill(j))
\end{eqnarray*}

\subsection{Reaching Definitions Analysis}
\label{subsubsec:sa-reaching-definitions}

The goal of this analysis is to determine the set of statements that create (i.e., define) variable values that may reach the current statement. This information can be checked to determine, for example, if there can be any reaching definitions within the current loop for variables involved in the current statement. A statement with no such definition can be hoisted outside the loop.

Similar to alias analysis, this is also a forward dataflow analysis. $gen(j)$ consists of statements that define variable values. Any such statement also kills all reaching definitions for the variables being assigned to in this statement. $kill(j)$ is the set of statements whose definitions were killed in this basic block. The analysis equations are given below. 
\begin{eqnarray*}
reach\_in(j) &=& \bigcup_{i \in pred(j)} reach\_out(i)\\
reach\_out(j) &=& gen(j)\bigcup\,(reach\_in(j)-kill(j))
\end{eqnarray*}


\subsection{Type Inferencing}
\label{sec:type-inferencing}

Type inferencing for program variables is crucial for automatically generating C++ code that can be compiled and executed much faster than interpreting the given program. In this subsection we give a brief overview of datatypes in {\R} and our type inferencing rules.



Types in {\R} are either basic (atomic) or constructed from basic types.
\begin{flalign*}
T ::=& T_B & \# \text{basic types}& \\
	 & T_C & \# \text{constructed types}&
\end{flalign*}

\subsubsection{Basic Types}
\label{subsec:types-basic}


\begin{flalign*}
T_B ::=& NULL &	&										\\
	   & T_{NB} & \# \text{Non-NULL basic types}&		
\end{flalign*}
\begin{flalign*}
T_{NB} ::=& expr	& \# e.g., 1+2,x-y &		\\
	   & T_{NBE} & \# \text{Non-\{NULL$|$expr\} basic types}&		
\end{flalign*}
\begin{flalign*}
T_{NBE} ::=& raw & \# e.g., 00, 02 &			\\
   	   & logical & \# TRUE, FALSE, NA &			\\
	   & integer & \# \mathbb{Z}\quad e.g., 1L, 2L &			\\
	   & double & \# \mathbb{R}\quad e.g., 1, 2.3 &			\\
	   & complex & \# \mathbb{C}\quad e.g., 0+1i &				\\
	   & string	& \# e.g., `c',`abc' &		
\end{flalign*}

In {\R}, the $double$ data type is called ``numeric'' and the $string$ data type is called ``character''. The value $NA$ (Not Available) is a logical constant, but it has an equivalent representation for the other $T_{NBE}$ types, except $raw$, e.g., $NA\_integer\_$, $NA\_complex\_$, etc.

\subsubsection{Constructed Types}
\label{subsec:types-constructed}


For the following, we use the notation $\langle \;\; \rangle$ to denote a sequence. We use ``$\ldots$''' as a convenience to avoid expanding the full notation.

\vspace{5pt}
\bgroup
\def\arraystretch{1.4}
\noindent\begin{tabular}{>{$}l<{$}>{$}l<{$}>{$}r<{$}}
T_C ::=& vector(t) & t\in \{T_{NB} \cup list\}\\
       & list(\langle t \rangle) & t\in T\\
       & factor(\langle integer \rangle,\langle string \rangle) &\\
       & matrix(t) & t\in \{T_{NB} \cup list\}\\
       & \multicolumn{2}{>{$}l<{$}}{dataframe(\langle vector(t)|factor(\ldots)\rangle)}\\
       & & t\in \{T_{NBE} \cup list\}\\
       & array(t) & t\in \{T_{NB} \cup list\}\\
       & function(list(...),t_r) & t_r\in T
\end{tabular}
\egroup
\\

A $vector$ is a sequence of elements all of the same type. It is a datatype with a single dimension, the vector length, which is equal to the number of elements in the sequence. An $array$ is a datatype with multiple ($\geq$ 1) dimensions. A $matrix$ is a datatype with two dimensions. Like vectors, all elements of any array or matrix must be of the same type. On the other hand, a $list$ is a sequence of elements where each element can be of a different type.

A $dataframe$ is a sequence of vectors of the same type. It is a datatype with two dimensions, and all elements in the same column have the same type since they they belong to the same vector. In addition, all column vectors are of the same length. A key difference between a dataframe and a matrix is that \textit{all} elements in the entire matrix are of the same type. 

A $factor$ is a datatype that is commonly used to represent categorical data. It uses two sequences to correspond to a given sequence of (categorical) data elements. The first sequence of the factor represents the data elements as a sequence of integers, with each value being the position of the corresponding element in the second sequence (called levels). The levels of the factor consists of the unique elements, represented as strings, in the given data elements.

\subsubsection{Attributes}
\label{subsec:types-attributes}


Objects in {\R} have \textit{attributes} in addition to data. These track additional information beyond the type of the object. For example, \textit{dimensions} tracks the number of dimensions of arrays, \textit{dim} in an integer vector with each element tracking the size of the corresponding dimension of matrices, arrays or dataframes, \textit{nrow} and \textit{ncol} are integers that track the number of rows and columns respectively while \textit{rownames} and \textit{colnames} are string vectors that track their names. For objects (e.g., vectors, lists), \textit{length} (technically not an attribute in {\R}) tracks the number of elements while \textit{names} tracks their names. The \textit{class} attribute supports object-oriented programming by tracking the class whose methods need to be invoked. Users can also add their own attributes. 

Attribute values need to satisfy some constraints depending on the data type. For example, the product of \textit{nrow} and \textit{ncol} must equal the number of elements, etc. Attributes values can be overwritten by the user in which case the object may be resized with $NA$ filled in for missing values.

For our example programs, checking for the potential use of the \textit{rownames} attribute is most useful as it can avoid performing costly string conversions at run-time as discussed in Section~\ref{subsec:ineff-attrib} in the context of the Kmeans example.

\subsubsection{Type Rules}
\label{subsec:types-subtyping}

Every variable having a basic type is a 1-element vector of that type. For brevity, we will sometimes use $vector(t)$ at places where a single element of type $t$ is expected.

\begin{table}[ht]
\centering\vspace{-10pt}
\caption{Subset of Type Inference Rules}
\resizebox{0.45\textwidth}{!}{
\begin{minipage}{0.45\textwidth}
\begin{center}
\begin{tabular}{|@{}c@{ }@{ }c@{}|}\hline
{\bf R1}& \inferrule{a_1:T_{a1} \\ \dots \\ a_k : T_{ak}}{c(a_1,\dots,a_k): \theta(\vee_{i=1}^{k} T_{ai}), \theta\in\{vector, list\}}\\[11pt]
{\bf R2}& \inferrule{a:T_a \\ b:T_b \\ OP \in \{+,-,*\}}{a \; OP \; b : T_a \vee T_b}\\[11pt]
{\bf R3}& \inferrule{a:T_a \\ b:T_b}{a \mathbin{\char`\^} b : vector(double)}\\[11pt]
{\bf R4}& \inferrule{a:T_a \\ b:T_b \\ OP \in \{=, {<}{-}, {<}{<}{-}\}}{a \; OP \; b : T_b}\\[11pt]
{\bf R5}& \inferrule{a: T_a \\ b:T_b \\ OP \in \{{[}{<}{-}, {[}{<}{<}{-}\}}{a \; OP \; b : T_a \vee T_b}\\[11pt]
{\bf R6}& \inferrule{a:T_a}{a[\dots]: \tau(c(a))}\\[11pt]
{\bf R7}& \inferrule{a:T_a \\ b:T_b}{a{:}b : \tau(c(a))}\\[11pt]
{\bf R8}& \inferrule{a:T_a \\ t \in \{logical, integer, double\}}{as.t(a):vector(t)}\\[12pt]
{\bf R9}& \inferrule{a:T_a \\ b:T_b \\ OP \in \{numeric, rnorm, rbeta, runif, sqrt, floor, {\char`\\}\}}{a \; OP \; b : vector(double)}\\[11pt]
{\bf R10}& \inferrule{a:T_a \\ b:T_b \\ OP \in \{length, nrow, ncol\}}{a \; OP \; b : vector(integer)}\\[11pt]
{\bf R11}& \inferrule{a:T_a \\ OP \in\{sample, rep\}}{OP(a,\dots) : \tau(c(a))}\\[11pt]
{\bf R12}& \inferrule{ }{paste(\dots):vector(string)}\\[11pt]
{\bf R13}& \inferrule{a:T_a \\ b:T_b \\ c:T_c}{if(a)\; b \;else\; c : T_b \vee T_c}\\[11pt]
{\bf R14}& \inferrule{a:T_a}{return\; a: T_a}\\[11pt]
{\bf R15}& \inferrule{a_1:T_{a1} \\ \dots \\ a_k : T_{ak}}{\{a_1 \dots a_k\}: T_{ak}}\\\hline
\end{tabular}
\end{center}
\end{minipage}
}\vspace{-5pt}
\label{tab:type-rules}
\end{table}
Table~\ref{tab:type-rules} shows a subset of type rules for various operations relevant to our examples. Each rule shows a horizontal line that separates the premises (above) from the conclusions (below). We use the notation $x: T$ to indicate that $x$ has type $T$. The notation $\tau(x)$ also denotes the type of $x$.

Rule R1 in Table~\ref{tab:type-rules} deals with the combine ($c$) operator that creates vectors or lists after coercing constituent elements to the same supertype. The subtyping relation for $c$ is:
\begin{math}
NULL \preceq raw \preceq logical \preceq integer \preceq double \preceq complex \preceq string \preceq list \preceq expr
\end{math}.
This can induce subtyping in a natural manner on complex types~\cite{pierce:types-programming-languages:book:2002}.

We use the notation $\vee$ to denote a type join. So, for example, $\vee(integer,string)=string$, $\vee(vector(integer),string)$ $=\vee(vector(integer),vector(string))=vector(string)$, etc. Rule R1 says that the $c$ operator performs a type join on the inputs and returns a $vector$ or $list$.

The following example illustrates this rule and the subtyping relation described above. \texttt{x} is a vector constructed from an $integer$ (1L), $logical$ (FALSE), $double$ (2.3), $string$ (``a''), and $complex$ (2+3i). \texttt{str(x)} shows that \texttt{x} is of type $string$ and all its elements have been coerced to this type.
\begin{lstlisting}[frame=single,frameround=tttt]
> x <- c(1L,FALSE,2.3,"a",2+3i)
> str(x)      #returns object structure
 chr [1:5] "1" "FALSE" "2.3" "a" "2+3i"
\end{lstlisting}


Rules R2 and R3 deal with arithmetic operations with the same subtyping relation as above, but restricted to $logical$, $integer$, $double$ and $complex$.
Rule R4 deals with full assignment whereas R5 describes sub-assignment where the type of the LHS can change. For example, an assignment to an element of a $vector$, $matrix$, or $array$ will result in the coercion of all constituent elements to the supertype.
Rule R7 describes the range operator ($:$). Rules R8--R12 describe the results of various utility functions. Rules R13--R15 deal with control flow and code blocks.


To infer types, one needs to apply the rules presented in Table~\ref{tab:type-rules} to each operation in the program repeatedly till a fixed point is reached. We currently do inferencing only for a subset of the types in {\R}. These are $logical$, $integer$, $double$, $complex$, $string$, $vector$, $matrix$, $array$ and $function$ types. 

For code translation, we make a simplification to Rule R9 in Table~\ref{tab:type-rules} for the $floor$ function. While its return type is $double$ in {\R}, we consider it as $integer$ so that more efficient code can be generated. Thus, variable $mid$ in the Binary Search example (Listing~\ref{lst:binsrch}) gets assigned type $integer$.

\subsection{Other Analyses}
\label{subsubsec:sa-other}

Here we list a few other important analyses that are useful for optimizing {\R} programs.

\textbf{Loop Invariants Analysis}: This determines which computations are invariant across all iterations of the enclosing loop and then moves them out of the loop. This saves runtime by avoiding repeated evaluation of invariant computations. The Unique Genotypes Test program (Listing~\ref{lst:ugt}) can benefit from this analysis since allocations of \texttt{genos} and \texttt{genos.c} can be hoisted out of the loop. We also discuss this analysis in Section~\ref{subsec:optims-vectorization} in the context of vectorization.

\textbf{Loop Analysis}: This determines loop properties such as number of iterations. This enables identification of vectorization opportunities. 
Other properties such as data dependence and reuse help to identify loop tiling opportunities, e.g., in the Matrix Multiplication program (Listing~\ref{lst:matmul}).

\textbf{Array Index Analysis}: This determines the indexing values for array accesses. This analysis can be helpful to other analyses as otherwise, the entire array will be treated as one object leading to conservative information and lost opportunities for optimization.

\subsection{R-specific Challenges and Limitations}
\label{subsec:sa-limitations}

The {\R} parser identifies entire control structures (\texttt{if}, \texttt{for}, \texttt{while}) as single S-expressions that includes the bodies of the constructs as other expressions. During CFG construction, S-expressions need to be carefully broken into lists of simple statements with each statement represented by a node.

{\R} has the super-assignment operator ({<}<-) that allows assignment to a variable in the enclosing environment (e.g., parent function). Fortunately, {\R} is lexically scoped, so variable names can be resolved statically provided that the code is not modified dynamically.

{\R} programs can overwrite code on-the-fly and modify the execution environment. This invalidates program facts determined by prior static analysis of the code. Disallowing code modifications and having sealed environments~\cite{tierney:byte-code-compiler:iowa:2014} can avoid these situations.


\section{Optimizations}
\label{sec:optims}

We now describe how the statically inferred program properties are used to optimize evaluation of a given {\R} program. We discuss three optimizations to illustrate the concepts.

\subsection{Space Reuse}
\label{subsec:optims-space-reuse}

The goal of this optimization is to avoid unnecessary memory allocations in the {\R} interpreter by reusing existing space allocated to variables. The guiding principle is that allocated space can be reused if there are no live aliases and the variable is not live beyond this point (value not needed again before re-definition). The live variable and alias analyses provide the information necessary for this optimization. 

We augment the S-expression structure with the following fields to enable this optimization.

\quad{\bf Dead-variable pointers, Not-live flag}: The pointers identify variables in the RHS of a computation that are dead (not live) after the computation. The Not-live flag is set in the S-expressions of each of those variables to indicate that their space can be reused.

\quad{\bf No-alias flag}: This flag is set if the destination variable of the computation is not aliased, or its aliases are not live. It indicates that a copy is not needed on a write to the variable.

\quad{\bf Original vector length}: This flag is set during in-place resizing of vectors. We need this flag so that book-keeping operations by the garbage collector are not affected. We perform in-place resizing only when the target vector length is less than the original vector length.

The {\R} interpreter already uses macros MAYBE\_SHARED, defined as a check for the $named$ field value to be greater than 1, and NO\_REFERENCES, defined as a check for the $named$ field value to be equal to 0, during interpretation to check if copies should be made. We augment these macros to include information about program facts. 

Functions involved in interpretation also need modification to process the additional information. For example, the main interpretation function, \texttt{eval}, checks variables against dead variable information for that statement and sets flags appropriately. Operator implementation functions need to check for the possibility of reuse before new allocations. 

Consider the Copy-On-Write program (Listing~\ref{lst:cow}). The basic data type in this program, $double$, occupies 8 bytes. Thus, Line 2 allocates 8n bytes. Line 3 causes \texttt{x} and \texttt{y} to be aliased. By default, Lines 4 and 5 cause new allocations for copies requiring a maximum of 24n bytes of temporary allocation (for large n) and 16n bytes of steady-state allocation (for \texttt{x} and \texttt{y}). However, neither \texttt{x[2]} nor \texttt{y[2]} have live aliases, so the modifications can be done in place leading to 16n bytes reduction in temporary allocation and 8n bytes of steady-state allocation. Some optimization opportunities may be lost if the analysis cannot distinguish between different elements of the vectors leading to entire arrays being treated as a single object. 

Next, consider the Simple Arithmetic program (Listing~\ref{lst:sa}). Lines 4 and 5 each result in 8n bytes of steady-state allocation. By default, Line 6 causes an additional 8n bytes allocation for \texttt{d} leading to a total of 24n bytes of steady-state allocation. However, neither \texttt{x} nor \texttt{y} are used beyond the computation in Line 6, so their space can be reused by the subtraction operator function and the modifications can be done in-place. This method avoids allocation of temporary vectors of size 8n each for computations \texttt{(x-xs)} and \texttt{(y-ys)}. This approach also leads to savings in steady-state allocation since \texttt{d} can use the space from the computations, resulting in a total of 16n bytes of steady-state allocation.

To apply this optimization, we need to identify the last use of source variables in the functions implementing various operations, and create a mechanism for in-place computations. The latter is not always easy. For example, in the Matrix Multiplication program (Listing~\ref{lst:matmul}) both \texttt{A} and \texttt{B} are dead beyond the multiplication, but the default multiplication algorithm always allocates new space for the result.

We also need to add fields to the {\R} S-Expression structure. These changes increase the size of every S-Expression object, thereby needing more memory and potentially reducing cache performance. We maintain two copies of the {\R} interpreter, the unmodified one and one with the augmented structures. The augmented interpreter is chosen only if the space reuse optimization is applied.

\subsection{Vectorization + Code Motion}
\label{subsec:optims-vectorization}

The goal of this optimization is to identify program statements that can be replaced with efficient vector equivalents and move such statements outside enclosing loop(s) if necessary. The reaching definitions analysis provides the information necessary for simple applications of this optimization. More advanced applications also need information about variable types from type inferencing.

Consider the Simple Vectorization program (Listing~\ref{lst:vectorization}). The computation on Line 5 adds corresponding elements vectors. If the number of iterations matches the object lengths, then {\R} allows specifying this entire sequence of operation with a single statement, \texttt{z = x + y}. This statement is more efficient since the looping over the elements is implemented internally and does not require interpretation. Moreover, the statement \texttt{z = x + y} can be moved outside the loop as otherwise the same computations would be repeated on every iteration of the loop.

In general, a computation is loop invariant if it is a constant or all reaching definitions of all variables involved in the computation come from outside the loop or other loop-invariant computations. Normally, \texttt{z[i] = x[i] + y[i]} is not loop-invariant since the reaching definition of \texttt{i} is generated within the loop. However, after the vectorization transformation this is no longer the case, and the computation becomes loop invariant.

More sophisticated applications of this transformation require type information. For example, consider the computation \texttt{t=t+(X[i]-Y[i])} within a loop. Vectorizing this code to \texttt{t=t+(X-Y)} is incorrect since it changes the type of \texttt{t} (e.g., from a basic type to vector type). Insertion of additional aggregation operations may be needed in such cases, e.g., \texttt{t=t+sum(X-Y)}. Currently we do not handle this case.

\subsection{C++ Code translation}
\label{subsec:optims-codegen}

The goal of this optimization is to translate the {\R} program (or a portion of it) to C++, which is then compiled into a shared library, loaded by the {\R} interpreter, and directly executed without further interpretation. Type inferencing provides the information necessary for this optimization. 

We show two examples of the generated code below to illustrate some features. Listing~\ref{lst:codegen_random_walk} shows a portion of the translated code for the 2D Random Walk program. The outermost function uses {\R} S-Expressions (SEXPs) as parameters and also return SEXPs. This enables the translated code to interface with {\R} for receiving inputs and returning results. The function is called through {\R}'s \texttt{.Call} interface.

We represent constructed types using C++ objects, e.g. \texttt{Vector}. The \texttt{r2c} and \texttt{c2r} functions convert between {\R} and C++ representations. Conditional assignment may require some code transformation. Line 5 of the {\R} code in Listing~\ref{lst:2dwalk} gets translated to Lines 11--15 in the C++ code with the assignment to \texttt{delta} placed in both the \texttt{if} and \texttt{else} code blocks. Another transformation involves replacing calls to {\R} functions with C++ versions (e.g., \texttt{c\_numeric}) for a subset of functions and types. In other cases, functions provided by {\R} are called through our \texttt{r\_internal} function that calls {\R}'s \texttt{eval} function to invoke and interpret the required function.

Automatic code translation involves some other challenges that our system currently does not handle. This includes renaming for variables that have context-dependent types, handling nested function definitions, and recycling of values for automatically extending the length of variables.

\begin{lstlisting}[frame=single,frameround=tttt,caption={C++ code for 2D Random Walk},label=lst:codegen_random_walk]
SEXP rw2d1(SEXP r_n, SEXP rho) {
  int delta;
  int i;
  Vector<double> xpos;
  Vector<double> ypos;
  ext_prepare(rho);
  int n = r2c_int(r_n);
  xpos = c_numeric(n);
  ypos = c_numeric(n);
  for(i=2;i<=n;i++) {
   if (c_gt(r2c_Vector_double_(r_internal("runif", 1, c2r_int(1))), 0.5)) {
     delta = 1;
   } else {
     delta = -(1);
   }
   ...
  }
  SEXP r_ret = r_internal("list", 2, c2r_Vector_double_(xpos), c2r_Vector_double_(ypos));
  ext_finalize();
  return(r_ret);  
}
\end{lstlisting}


\begin{figure*}[ht]
\centering
\subfloat[Gains with optimizations implemented]{\includegraphics[width=0.48\textwidth]{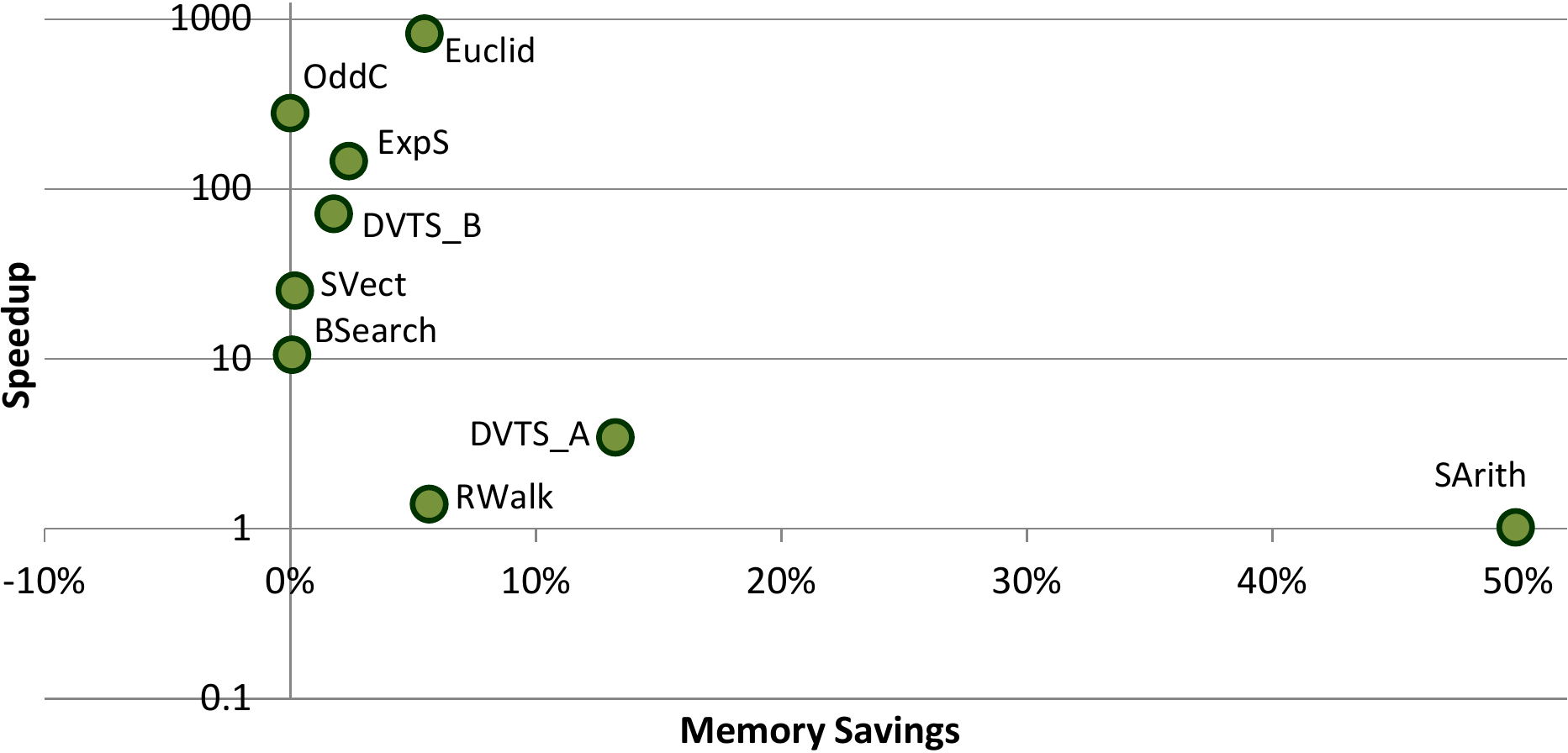}\label{fig:results_automated}}\vspace{4pt}
\subfloat[Gains with optimizations currently requiring user input]{\includegraphics[width=0.48\textwidth]{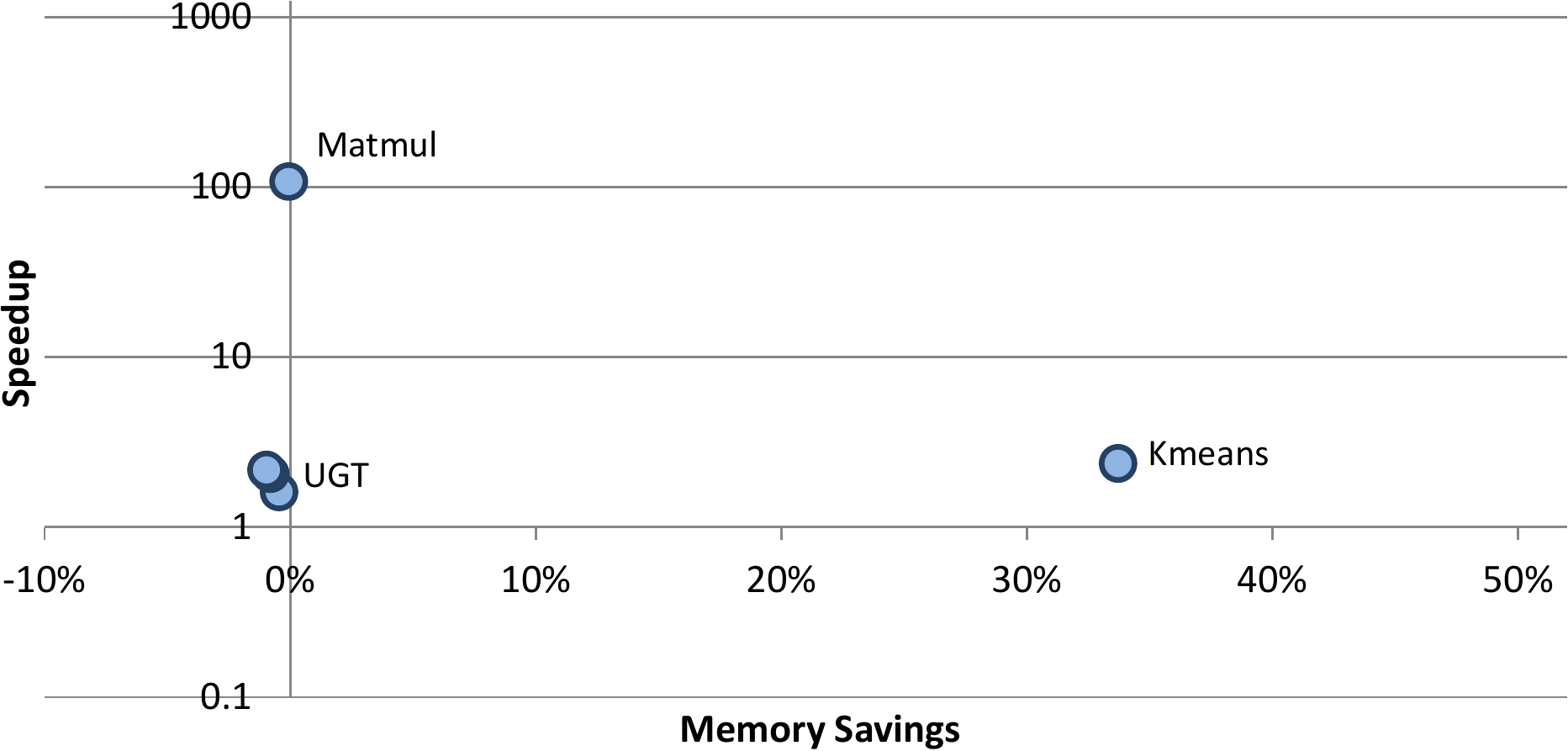}\label{fig:results_expected}}
\caption{ROSA improvements compared to CRAN {\R}.}
\label{fig:results}\vspace{-11pt}
\end{figure*}
\begin{figure*}[ht]
\centering
\subfloat[Gains with optimizations implemented]{\includegraphics[width=0.48\textwidth]{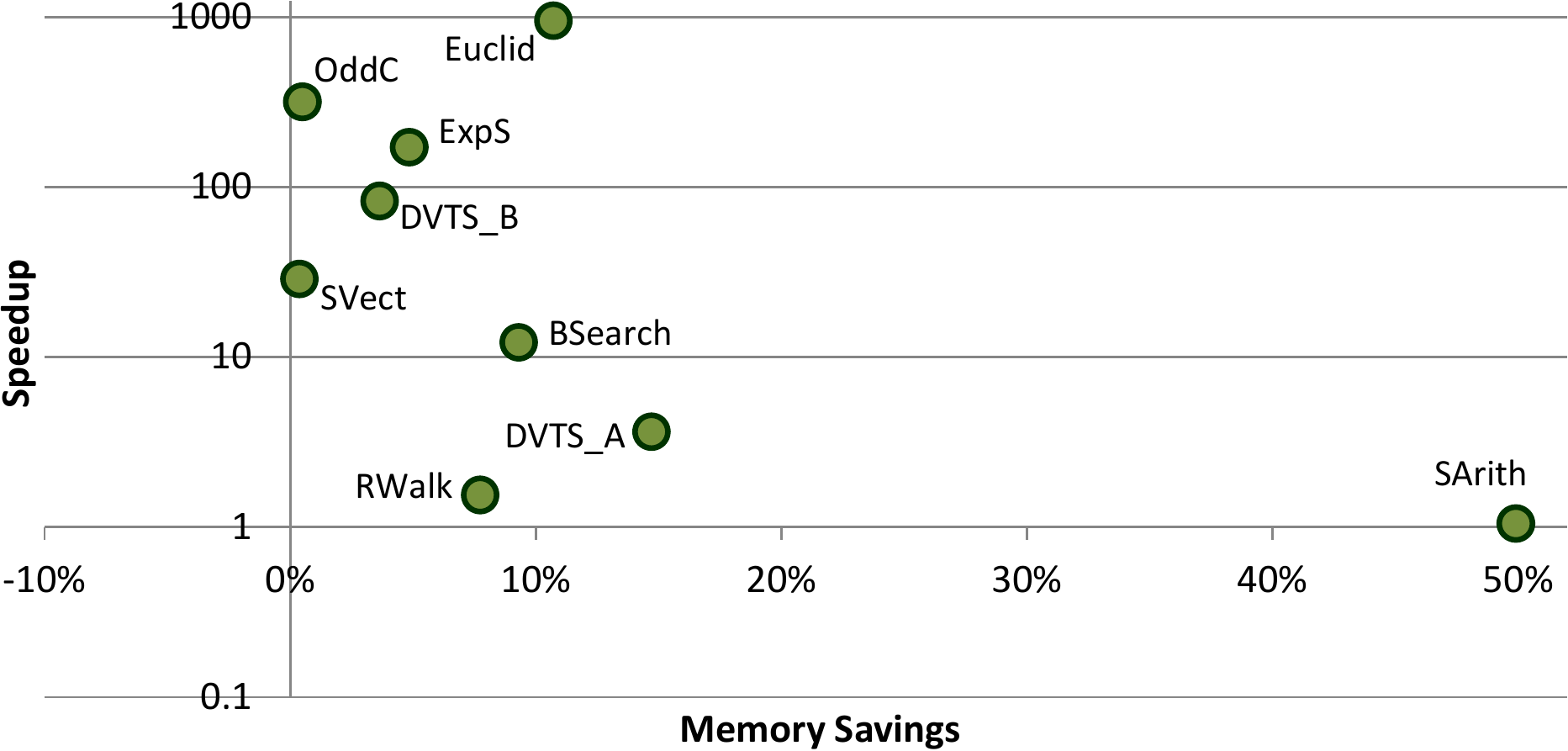}\label{fig:results_automated_MRO}}\vspace{4pt}
\subfloat[Gains with optimizations currently requiring user input]{\includegraphics[width=0.48\textwidth]{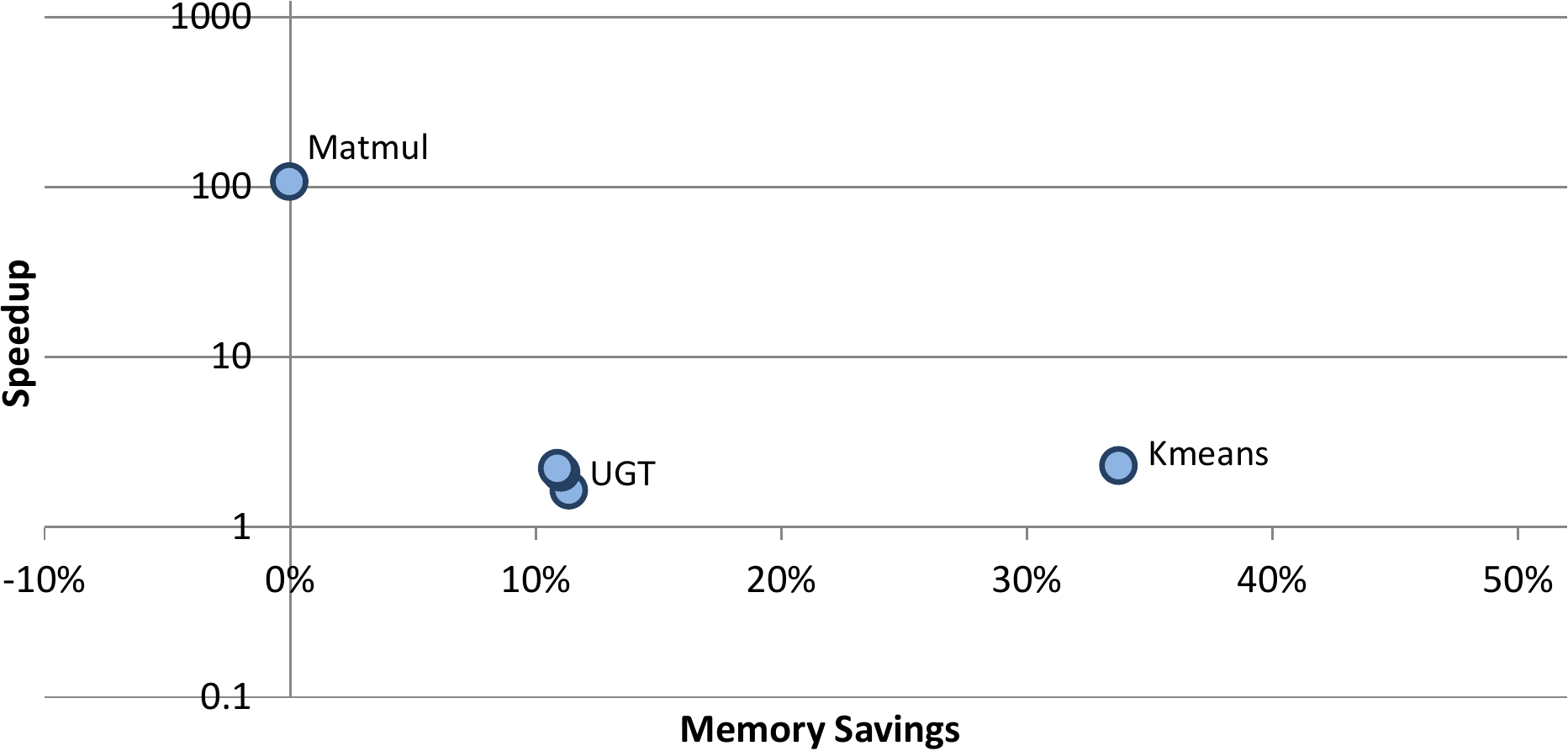}\label{fig:results_expected_MRO}}
\caption{ROSA improvements compared to MRO.}
\label{fig:results_MRO}
\end{figure*}

\section{Experiments}
\label{sec:experiments}

In this section we evaluate the effectiveness of ROSA in reducing execution time and memory footprint requirements for evaluating {\R} programs. We design our experiments to answer the following questions.
\begin{enumerate}
\item How well does ROSA improve upon the standard open-source version of R (CRAN R)? We answer this question in Section~\ref{subsec:experiments-improvements-baseline}.
\item How well does ROSA improve upon prior works that also transparently improve upon standard R interpretation? We answer this question by comparing ROSA with two works---Microsoft R Open (Section~\ref{subsec:experiments-improvements-MRO}) and R Byte Code Compiler (Section~\ref{subsec:experiments-improvements-bcc}).
\end{enumerate}

\subsection{Workloads and Setup}
\label{subsec:experiments-workloads}

As our workload we used the {\R} programs shown in Listings 1, 3--7, 9--14. These programs were used as workloads in prior works on optimizing {\R}~\cite{templelang:advanced-compilation-tools:ss:2014}, profiling {\R}~\cite{sridharan:profiling-contemporary-processor:vldb:2014}, {\R} applications~\cite{reynolds:ugt-r-script:2011,reynolds:genetic-equilibrium-conservation:phdthesis:2011,zhang:riot-i/o-efficient:cidr:2009}, and are key examples in books on {\R}~\cite{matloff:art-programming:book:2011,wickham:advanced-r:book:2014}. We have adapted some programs, e.g., to have larger input data sizes. We show the inputs for each program in their listings; e.g., the 2D Random Walk program, Listing~\ref{lst:2dwalk}, operates on 10M elements (\texttt{n <- 1e7}). The Kmeans program, Listing~\ref{lst:kmeans}, uses the Airline on-time dataset~\cite{airline-ontime-dataset:2009}.

We run our experiments on a 2.6 GHz dual-socket 
Intel Xeon E5-2660 v3 (Haswell) server machine with 25 MB of last-level cache, and 10 cores per socket, 
160 GB of main memory, and running Ubuntu 14.04.1 LTS. We measure execution time as follows: For programs that have \texttt{system.time(...)}, it is the elapsed time reported in the outputs of those statements. For other programs, it is the elapsed time for the entire program. We calculate speedup as (baseline time)/(new time). 
We determine memory usage by using the \texttt{getrusage} function in Linux and reading the value of the \texttt{maxrss} (maximum resident set size) field in the result. The memory savings for a workload execution is the reduction in maximum resident set size (RSS) for the entire program and is calculated as: $1 - \frac{\text{new max. RSS}}{\text{baseline max. RSS}}$.

\subsection{Improvements over CRAN R}
\label{subsec:experiments-improvements-baseline}

Figure~\ref{fig:results_automated} shows improvements, relative to CRAN {\R}-3.2.5~\cite{r:cran} (baseline), with optimizations that are currently automated in ROSA (i.e., those optimizations that are marked as `Automated' in Table~\ref{fig:architecture}). In the figure, we plot each program as a point/bubble and show the memory saving on the x-axis and performance improvement on the y-axis. 

As can be observed in  Figure~\ref{fig:results_automated}, ROSA saved \myapprox50\% memory for the Simple Arithmetic program through the space reuse optimization (Section ~\ref{subsec:optims-space-reuse}). It sped up the Simple Vectorization program by \myapprox25$\times$ through the vectorization and code motion transformations (Section~\ref{subsec:optims-vectorization}). 

C++ code was generated (Section~\ref{subsec:optims-codegen}) for the remaining programs. All of these, except 2D Random Walk showed large speedups. Euclidean Distance, with its three level nested loops, showed the most speedup of nearly three orders of magnitude over the baseline execution. 2D Random Walk showed a smaller, but still significant performance improvement of \myapprox40\%. It incurred overheads due to repeatedly calling the \texttt{runif(1)} function for execution within the {\R} interpreter. Vectorizing this call should provide further speedups. Memory savings for compiled code were achieved by reducing/eliminating interpretive overhead and associated creation of temporary objects, and having C++ versions of data structure allocations.

Figure~\ref{fig:results_expected} shows improvements, relative to the baseline, with optimizations that require user input; i.e., those optimizations that are marked as `User-Input' in Table~\ref{fig:architecture}. ROSA currently needs additional information from the user and confirmation that the optimized program has the desired behavior. For example, the implementation of the matrix multiplication operator, \texttt{\%*\%}, is internal to the {\R} interpreter and not readily available to the static analyzer. The code fragment for the implementation needs to be identified. Transforming the implementation to have a more cache-friendly access pattern, as discussed in Section~\ref{subsec:ineff-cache}, created a speedup of 108 over the original implementation.

Strength reduction transformations are currently not fully automated in our system. User input is needed to check and apply the transformations. Two programs benefit from this transformation---Kmeans and Unique Genotypes Test.

Kmeans suffers from string conversion overheads because the function \texttt{na.omit} causes some rows in the input to be omitted, as they have \texttt{NA} values, and triggers a recalculation of row names leading to the sequence of computations described in Section~\ref{subsec:ineff-attrib}. The function \texttt{as.matrix.data.frame} has an efficient path for matrix conversion that does not compute row names if the input data frame has them in the form \texttt{(NA,-n)}, which is the case with the original data frame \texttt{A}. The overheads are eliminated through a strength reduction of data frame to matrix type by calling the \texttt{as.matrix} function on the input table before the other computations in the program. Memory footprint is also reduced because the large number of string objects are not created.

The Unique Genotypes Test program also suffers from string conversion overheads as discussed in Section~\ref{subsec:ineff-conversion}. The elements 0 and 1 passed to the \texttt{sample} function on line 7 in Listing~\ref{lst:ugt} are floats, changing them to 0\texttt{L} and 1\texttt{L} causes them to be treated as integers improving performance by \myapprox61\%. Changing them to ``0'' and ``1'' causes them to be treated as strings. This avoids the conversion altogether and more than doubles the performance compared to the baseline. Further optimizations for this program are possible by determining that the repeated creations (and allocations) of \texttt{genos} and \texttt{genos.c} are not necessary for every loop iteration and can be hoisted out of the loop (code motion). This leads to an additional \myapprox11\% improvement in performance compared to baseline. This analysis requires keeping track of loop bounds and matrix dimensions that is not currently automated in our system. We show all three variants of this program together in Figure~\ref{fig:results_expected}.

\subsection{Improvements over MRO}
\label{subsec:experiments-improvements-MRO}

Microsoft R Open (MRO)~\cite{MRO}, formerly known as Revolution R Open, is an open-source distribution of R that enhances CRAN R by supporting multithreaded execution for BLAS (Basic Linear Algebra Subprograms)/LAPACK (Linear Algebra Package) math libraries. The current release of MRO is based on CRAN R 3.2.5, which is why we also use that version of CRAN R throughout this paper so that a fair comparison can be made. We run the prebuilt 64-bit distributions of MRO with Intel MKL (Math Kernel Library) that provides the BLAS/LAPACK functions. 

Parallel execution of select math functions in MRO+MKL results in significant speedups over CRAN R (baseline). On our 20-core Haswell server we observe the following speedups (calculated as $= \frac{\text{CRAN R execution time}}{\text{MRO execution time}}$) for MRO performance benchmarks~\cite{MRO-benchmarks}: \myapprox162 (matrix crossproduct), \myapprox109 (Cholesky factorization), 1.03 (QR decomposition), \myapprox29 (singular value decomposition), \myapprox16 (principal component analysis), and 4.56 (linear discriminant analysis).

ROSA's improvements are not subsumed by MRO. Figures~\ref{fig:results_automated_MRO} and~\ref{fig:results_expected_MRO} show ROSA's improvements with respect to MRO (baseline) for optimizations that ROSA currently automates and for those that require user inputs. ROSA's improvements over CRAN R carry over to MRO as well. ROSA focuses on optimizing single-threaded executions of R code and its techniques are orthogonal to, and can be used in conjunction with, optimizations such as parallelizing math libraries. Like CRAN R, MRO also fails to run to completion for the Simple Arithmetic program (Listing~\ref{lst:sa}) when run with 9 billion elements. 

\subsection{Comparison with the byte code compiler}
\label{subsec:experiments-improvements-bcc}

Both CRAN {\R} and MRO include a byte code compiler~\cite{tierney:byte-code-compiler:iowa:2014}. Users can byte-compile functions in their programs into byte codes. Executing these compiled functions cause their evaluation by the \texttt{bcEval} function (instead of the \texttt{eval} function), which implements a byte code interpreter. Commonly-used constructs (e.g., \texttt{for}, \texttt{while}), arithmetic, and relational operators have optimized, inlined implementations (e.g., C calls for operations on scalar values) to speed up execution. Other expressions will be interpreted by the \texttt{eval} function.

\begin{figure*}[ht]
\centering
\subfloat[Compared to CRAN R + BCC]{\includegraphics[width=0.48\textwidth]{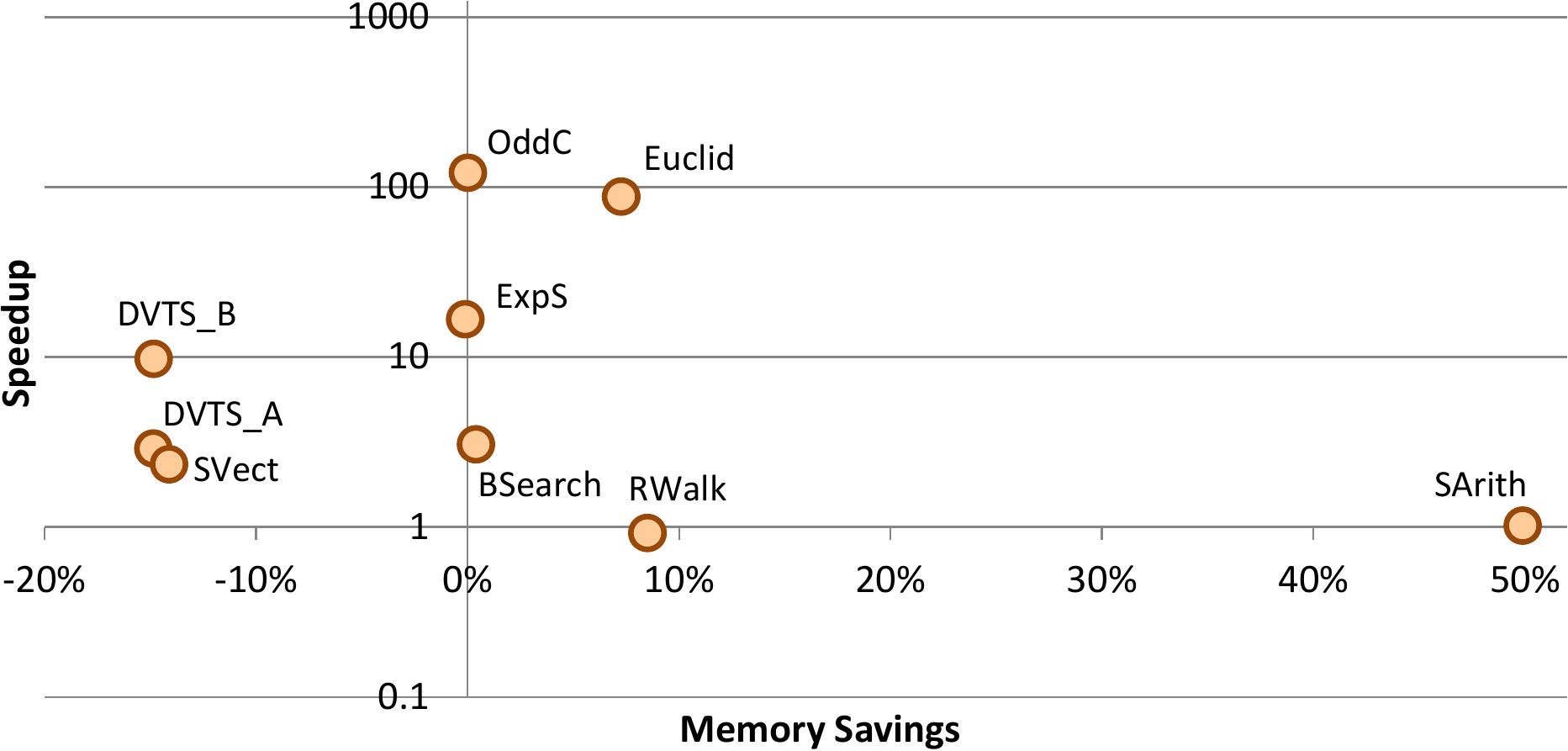}}\vspace{4pt}
\subfloat[Compared to MRO + BCC]{\includegraphics[width=0.48\textwidth]{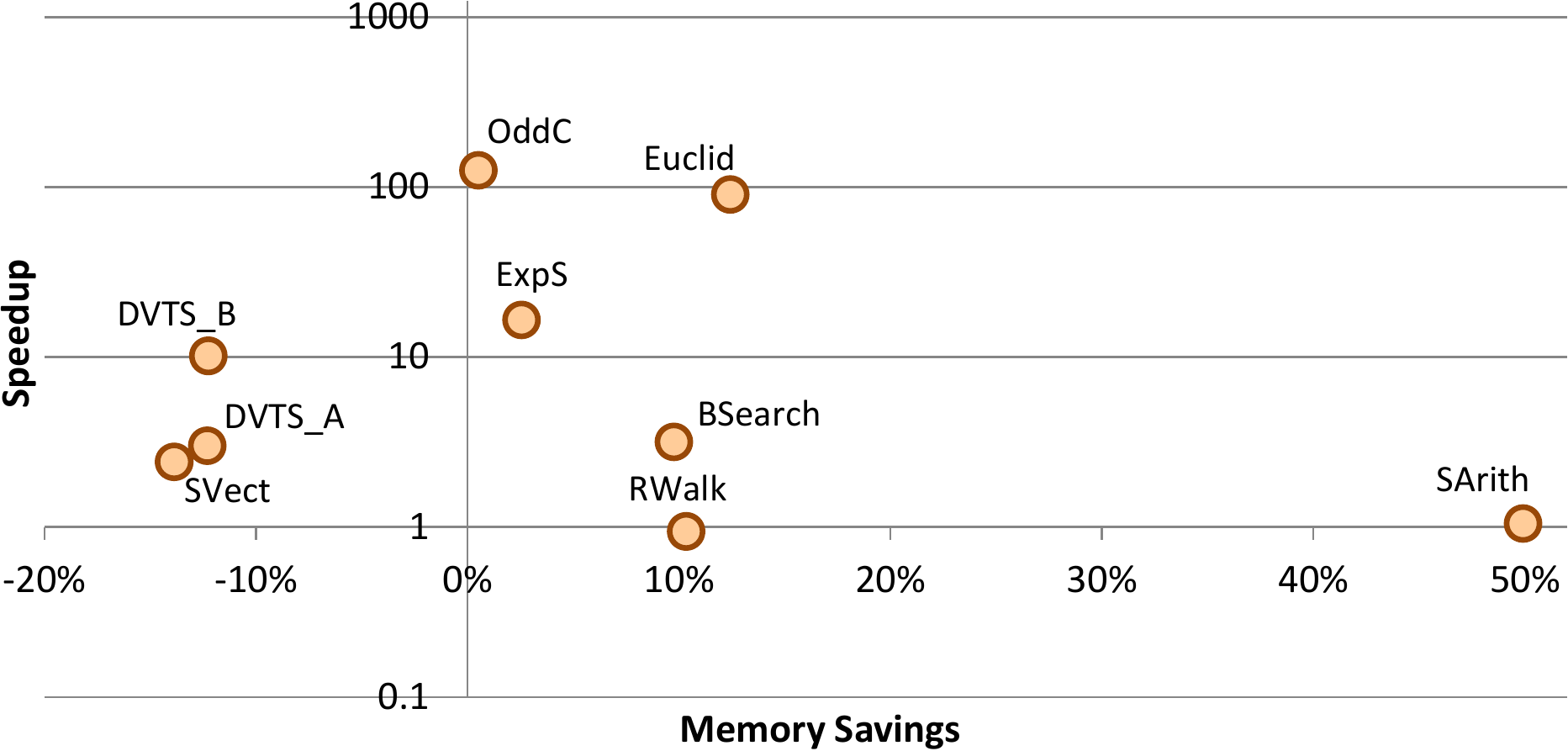}}
\caption{ROSA improvements, on optimizations it currently automates, relative to that by the byte code compiler (BCC).}
\label{fig:results-cmp}
\end{figure*}
Figure~\ref{fig:results-cmp} shows ROSA's improvements compared to the byte code compiler (BCC) with CRAN R and MRO. For these results, the execution with BCC is treated as the baseline. ROSA significantly outperformed BCC for many programs---e.g., by almost two orders of magnitude for Euclidean Distance. The gains are less compared to those over full interpretation (Figure~\ref{fig:results_automated}) as BCC somewhat optimizes execution. BCC uses {\R} objects and is interpreter-based, resulting in more overheads compared to execution of C++ code generated by ROSA. The only exception is the 2D Random Walk program where BCC slightly improved (7.6\%) performance over ROSA due to conservative assumptions by our current code generator.

BCC also does not do optimizations such as vectorization or space reuse. So, ROSA saves more execution time and memory footprint for the Simple Vectorization and Simple Arithmetic programs respectively. BCC saves more memory than ROSA for a few benchmarks, such as the Discrete Value Time Series programs, likely due to more efficient memory management of temporary objects for function returns.

\subsection{Impact and Overheads}
\label{subsec:experiments-summary}

Our evaluation confirms that in many cases, ROSA can significantly speed up {\R} programs and/or save memory, compared to both CRAN R and MRO. For optimizations it currently automates, ROSA also improves upon the {\R} byte code compiler in either performance, or memory footprint, or both. C++ code generation, vectorization, space reuse, strength reductions, and loop tiling transformations are important optimizations for {\R} program evaluations.

ROSA incurs overheads while performing static analysis of the input program and transforming it for optimized execution. However, these overheads are small. The main reason is that {\R} programs are usually short in length (number of lines of code) leading to small CFGs. For example, the CFG for the Euclidean Distance program has only 28 nodes. CFG construction for this program takes around 1 msec and static analysis takes approximately an additional 5 msec. In general, we expect that for most {\R} programs, the analysis will be completed within a few tens of msecs. A larger overhead arises from compiling generated code, with optimizations (-O2), and creating a shared library. This takes around 0.5 sec for our programs. Compilation overheads (5--25 msec) exist for BCC as well. Overall, the performance improvements far outweigh the overheads for original programs that are long-running (as in this paper) and/or repeated multiple times. We do not include analysis or compilation overheads while reporting speedups over baselines.


\section{Related Work}
\label{sec:related-work}

A number of prior works have explored techniques to improve {\R}'s efficiency, e.g., by optimizing the interpreter, using compiled C/C++ code, etc. Some of these approaches consider specific optimizations, e.g., improving vector operations~\cite{talbot:riposte-compiler-parallel:pact:2012}, C/C++ code translation~\cite{garvin:rcc-compiler:rice:2004,lang:rllvmcompile:git:2011,templelang:advanced-compilation-tools:ss:2014}. A number of approaches, e.g., FastR~\cite{kalibera:fast-ast-interpreter:vee:2014,r:fastr}, pqR~\cite{r:pqr}, Renjin~\cite{renjin}, Riposte~\cite{talbot:riposte-compiler-parallel:pact:2012}, etc. have developed new interpreters/evaluation engines for the {\R} language to reduce inefficiencies. 

A holistic framework for applying various analyses and optimizations, and which also works with the standard GNU {\R} interpreter is missing. ROSA fills this gap. Additionally, ROSA proposes a new type inferencing system for enabling automatic code translation. We study optimizations, such as space reuse and strength reduction, that enable processing of larger datasets, but have not been hitherto well explored.

{\R} includes interfaces that can be used to call external C and Fortran code. The external code can be compiled into shared libraries and loaded in {\R}. ROSA uses this interface to load and call compiled generated code. The Rcpp package~\cite{eddelbuettel:rcpp:JSSOBK:2011} allows integration of C++ code with {\R}. These packages provide interfaces and libraries, but do not automatically translate {\R} code of the user to C/C++/Fortran. The user has to translate/write the code manually. 

Garvin~\cite{garvin:rcc-compiler:rice:2004} developed RCC that compiles {\R} to C code. The generated code accesses the interpreter for object creation and incurs overheads due to time spent in variable definitions and lookups. In contrast, ROSA creates and manages C++ versions of data structures. This avoids the need to call the interpreter to manage or access variables. Type inferencing is crucial to enable this capability.

Temple Lang et al.~\cite{lang:rllvmcompile:git:2011,templelang:advanced-compilation-tools:ss:2014} proposed compiling {\R} code to LLVM IR. This can then be optimized by LLVM and re-targeted to different architectures. However, currently the user needs to provide the types of local variables and function signatures. Our proposed type inferencing system can address this issue to enable automatic compilation.

Tierney~\cite{tierney:byte-code-compiler:iowa:2014} and Wang et al.~\cite{wang:interpreter-level-specialization:cgo:2014} developed byte-code compilers that generate opcodes for a stack-based virtual machine. The byte code uses optimizations such as constant-folding and alternate function implementations to improve efficiency. In Section~\ref{subsec:experiments-improvements-bcc}, we showed that ROSA improved upon the byte code compiler freely available as part of the GNU {\R} distribution~\cite{r:cran}.

The RIOT system~\cite{zhang:riot-i/o-efficient:cidr:2009, zhang:riot-io-efficient:icde:2010} improves {\R}'s I/O efficiency by implementing an array storage manager and an optimization engine. Ricardo~\cite{das2010ricardo} and RHIPE~\cite{guha2012large} integrate {\R} with Hadoop to speed up processing by leveraging the inherent parallelism in data analysis workflows. These approaches are complementary to the techniques that we use in this paper.

Kotthaus et al.\cite{kotthaus:page-sharing-optimization:dls:2014} proposed a runtime optimization to reduce memory consumption by sharing memory pages and preventing unnecessary page allocations. For example, unmodified memory pages can be shared when objects are duplicated. In contrast, ROSA's space reuse optimization aims to save memory by preventing object duplication and allowing overwrites if aliased objects are no longer live. 

Although not the focus of this paper, we briefly mention Julia~\cite{julia}, a newer and high-performing dynamic programming language for scientific computing. Julia uses LLVM-based JIT compilation techniques to speed up evaluation. Julia's syntax is similar to, but not identical with, that of {\R}. However, a simple syntactic translation to convert {\R} programs to Julia and vice versa is not possible in general due to semantic differences between the languages. For example, if $x$ is an array, then the assignment $y=x$ has reference semantics in Julia whereas it has value semantics in {\R} with copy-on-write implementation policy. This means that modifications to elements of $y$ will be reflected in $x$ in Julia but not in {\R}. Another difference is that on an out-of-bounds assignment, arrays are automatically resized in {\R} whereas they are not in Julia. Unlike {\R}, Julia does not support lazy evaluation or the NULL type and has restrictions on logical indexing capabilities~\cite{julia-differences}. Currently, Julia's ecosystem seems to be less active compared to that of {\R}---at the time of this writing, there are 1437 registered packages for Julia~\cite{julia-packages} compared to 10962 for {\R}~\cite{cran-packages}, and Julia ranks significantly below {\R} in terms of popularity according to multiple indices~\cite{kdnuggets-ranking,tiobe-index-july-2016,redmonk-ranking,pypl-ranking}.


\section{Conclusions and Future Work}
\label{sec:conclude}

This paper presents ROSA, a framework for optimizing the evaluation of {\R} programs using static analysis techniques. These analysis techniques determine program facts that are then used to make {\R} programs execute more efficiently, either in terms of reduced execution time or reduced memory footprint. Such savings enable analysis of larger datasets on available hardware and within affordable run times while also leveraging the rich processing features of the {\R} language and computing environment. These savings are crucial in modern data platforms that increasingly package {\R} as a crucial component that in many cases runs {\R} code inside the data platform. Thus, ROSA extends the ability of such modern data platforms to use {\R} for analyzing large datasets.
Our future work will focus on expanding the set of analyses in ROSA and automating more transformations.

\section*{Acknowledgment}

This work was supported in part by grants from the National Science Foundation (NSF) under grants IIS-1250886 and by the National Institutes of Health (NIH) under grant U54AI117924.

\bibliographystyle{abbrv}
\bibliography{refs} 

\appendix
\section{Overhead associated with\\ Interpretation}
\label{sec:detailed-overhead-interpretation}

Figure~\ref{fig:subassign-flow} shows the detailed call sequence that is generated when interpreting the subassignment statement \texttt{y[2] <- 3} in the {\R} program that is shown in the upper right corner. One each line, the entries in blue show the main parameters passed to the function named on the left. We show only the main functions in the figure---other helper functions for symbol lookup and allocation of temporary variables are omitted. The statement includes two operators: the assignment operator (\texttt{<-}), implemented by the \texttt{do\_set} function, and the subassignment operator (\texttt{[<-}), implemented by the \texttt{do\_subassign} function. The {\R} interpreter creates a temporary variable `*tmp*' during interpretation of this statement. This temporary variable first points to the modified object, and finally \texttt{y} is set to point to the object. An optimized compiler, on the other hand, would implement the entire statement with a single memory access.

\begin{figure}[ht]
\centering
\fbox{\includegraphics[width=0.47\textwidth]{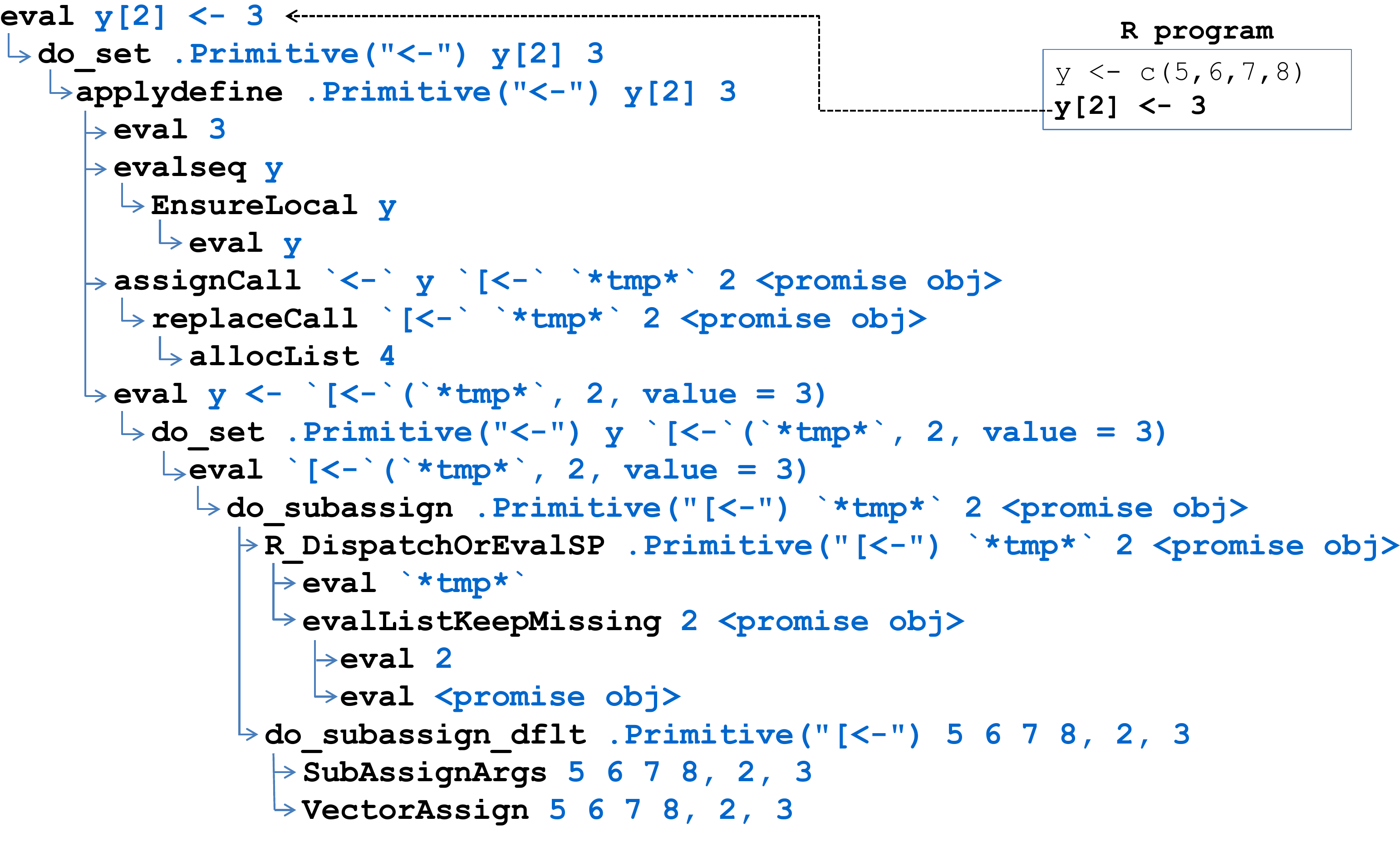}}
\caption{A part of the function call sequence during interpretation of subassignment statement.}
\label{fig:subassign-flow}
\end{figure}

Now consider Listing~\ref{lst:euclidean} and how simple type inferencing (followed by C/C++ code translation) can potentially help to eliminate the interpretation overhead. Line 5 implies that \texttt{ctr} is an integer and line 17 implies that \texttt{ans} is of type double or complex (a resolution can be made by analyzing whether or not the value of \texttt{total} is always $\geq0$). Since \texttt{nrow} and \texttt{ncol} return values of type integer, \texttt{nx}, \texttt{ny} and subsequently, \texttt{i}, \texttt{j}, \texttt{posX}, \texttt{posY} are inferred to be integers. \texttt{rnorm} returns double values, so \texttt{X} and \texttt{Y} are inferred to be constructed from type double.

\section{Overhead associated with\\ Attribute Evaluations}
\label{sec:detailed-overhead-attribute}

Consider the Kmeans program shown in Listing~\ref{lst:kmeans}. The \texttt{kmeans} function includes a statement \texttt{X <- as.matrix(X)} that converts argument \texttt{X} (which is a \textit{dataframe} object in this example) into a matrix. For this program, it also converts row numbers (integers) to row names (strings) to set the ``rownames'' attribute for \texttt{X}. These conversions are problematic if \texttt{X} has a large number of rows (\myapprox150M in this example) as costly integer-to-string conversions happen one-by-one for each row. This type conversion dominates processing time in the \texttt{kmeans} function.

\begin{figure}[ht]
\centering
\fbox{\includegraphics[width=0.47\textwidth]{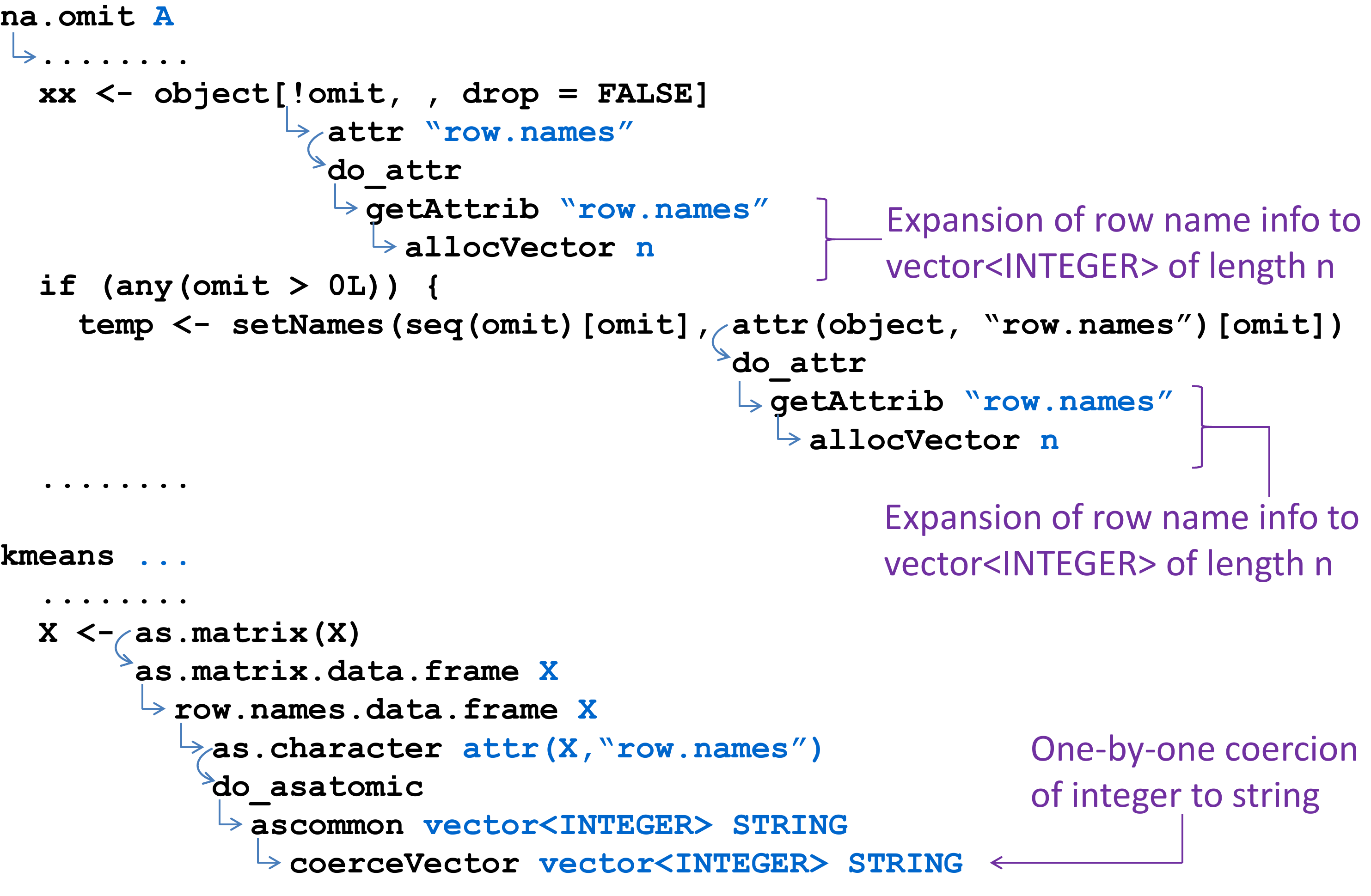}}
\caption{Type conversion during attribute computation.}
\label{fig:attrib-compute}
\end{figure}

Key portions of the detailed steps that are carried out by {\R} during the type conversion for the  \texttt{kmeans} program above is shown in Figure~\ref{fig:attrib-compute}. The internal {\R} matrix conversion function, called \texttt{as.matrix(...)}, is not always costly for dataframe objects. A dataframe is a list of constituent objects, each of which can be of a different type. Row name information for dataframes is usually maintained in compressed form ((\texttt{NA,-n}) for number of rows=\texttt{n}). This form prevents the type conversions, and the associated performance issue described above. However, the operation, \texttt{na.omit(...)}, expands the row name information while omitting rows that have \texttt{NA} (Not Available) entries.
Figure~\ref{fig:attrib-compute} shows a subset of the related call sequence in more detail. 
The expansion, along with allocation of large integer vectors of size \texttt{n}, happens during the processing of the \texttt{getAttrib} function. As Figure~\ref{fig:attrib-compute} shows, this processing and allocation happens twice. Next, when this object is passed to the \texttt{as.matrix(...)} function, the type conversion of integers to strings happens one-by-one for each integer.

\section{Example Analysis}
\label{sec:example-analysis}

Here we show part of the IR (Figure~\ref{fig:euclid-st}), CFG (Figure~\ref{fig:cfg-example}), and the inferred types of variables (Table~\ref{tab:types-euclidean}) for the Euclidean Distance program, Listing~\ref{lst:euclidean}.
\setlength{\DTbaselineskip}{5pt}
\DTsetlength{0.2em}{1em}{0.2em}{0.4pt}{1.6pt}
\begin{figure}[ht]
\centering
\begin{small}
\resizebox{!}{0.165\textheight}{
{\parbox{0.49\textwidth}{
\dirtree{%
.1 LANGSXP . 
.2 SYMSXP \hspace{5pt}\textcolor{blue}{for}.
.2 SYMSXP \hspace{5pt}\textcolor{blue}{k}.
.2 LANGSXP .
.3 SYMSXP \hspace{5pt}\textcolor{blue}{:}.
.3 REALSXP \hspace{5pt}\textcolor{blue}{1}.
.3 SYMSXP \hspace{5pt}\textcolor{blue}{p}.
.2 LANGSXP .
.3 SYMSXP \hspace{5pt}\textcolor{blue}{\{}.
.3 LANGSXP .
.4 SYMSXP \hspace{5pt}\textcolor{blue}{=}.
.4 SYMSXP \hspace{5pt}\textcolor{blue}{total}.
.4 LANGSXP .
.5 SYMSXP \hspace{5pt}\textcolor{blue}{+}.
.5 SYMSXP \hspace{5pt}\textcolor{blue}{total}.
.5 LANGSXP .
.6 SYMSXP \hspace{5pt}\textcolor{blue}{\^{}}.
.6 LANGSXP .
.7 SYMSXP \hspace{5pt}\textcolor{blue}{(}.
.7 LANGSXP .
.8 SYMSXP \hspace{5pt}\textcolor{blue}{-}.
.8 LANGSXP .
.9 SYMSXP \hspace{5pt}\textcolor{blue}{[}.
.9 SYMSXP \hspace{5pt}\textcolor{blue}{X}.
.9 SYMSXP \hspace{5pt}\textcolor{blue}{posX}.
.8 LANGSXP .
.9 SYMSXP \hspace{5pt}\textcolor{blue}{[}.
.9 SYMSXP \hspace{5pt}\textcolor{blue}{Y}.
.9 SYMSXP \hspace{5pt}\textcolor{blue}{posY}.
.6 REALSXP \hspace{5pt}\textcolor{blue}{2}.
}}}}
\end{small}
\caption{S-expression representation for lines 12--13 of the Euclidean Distance program. INTSXP, REALSXP, SYMSXP, and LANGSXP represent integers, reals, symbols and language structures respectively.}
\label{fig:euclid-st}
\end{figure}

\begin{figure}[!h]
\centering
\includegraphics[height=0.34\textwidth]{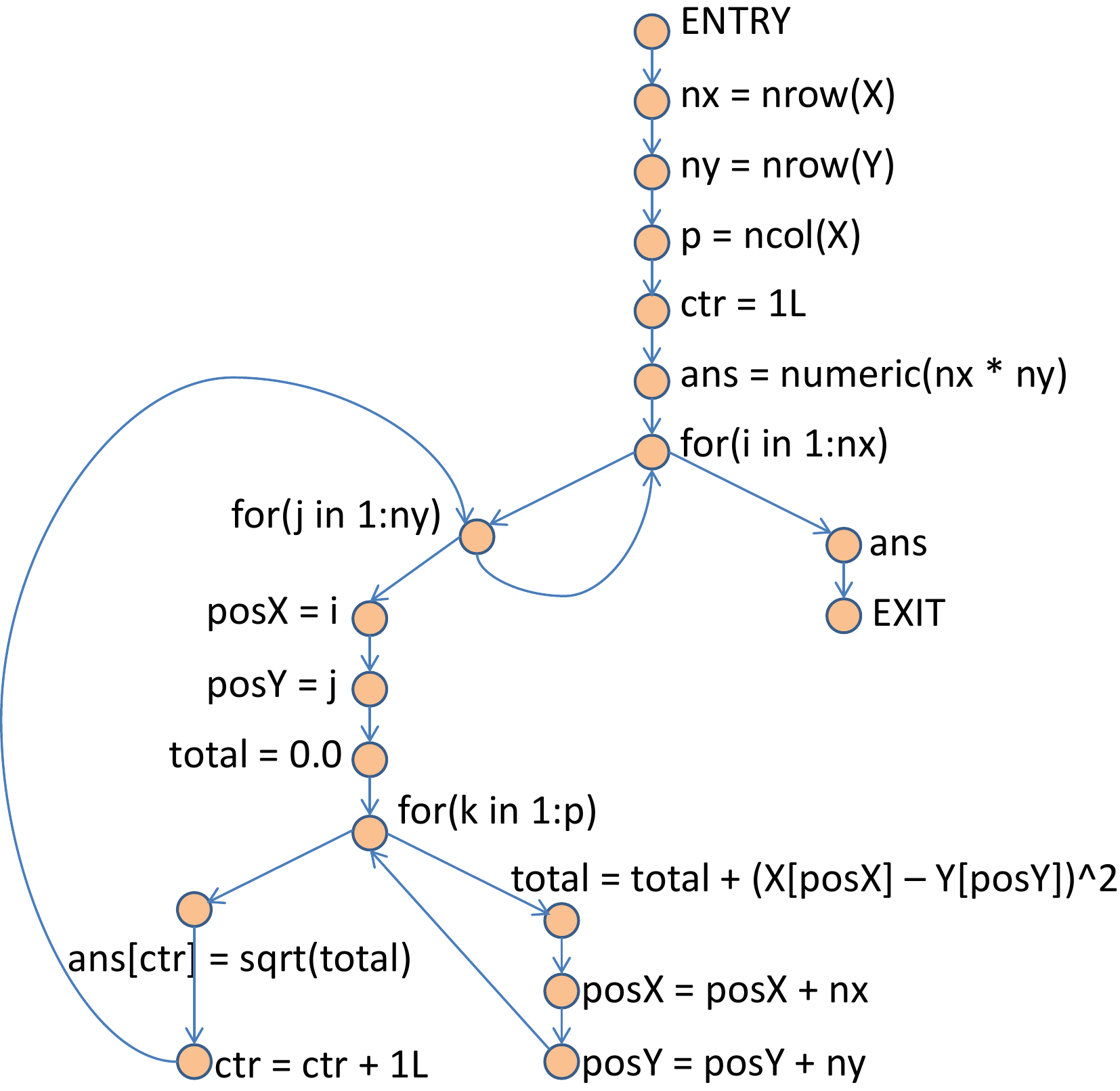}
\caption{CFG for function \texttt{dist} in Euclidean Distance program, Listing~\ref{lst:euclidean} (Appendix).}
\label{fig:cfg-example}
\end{figure}

\begin{table}[ht]
\centering
\caption{Variable types for Euclidean Distance}
\resizebox{0.42\textwidth}{!}{
\begin{tabular}{|@{ }c@{ }|c||@{ }c@{ }|c|}
\hline
\textbf{Variable} & \textbf{Type} & \textbf{Variable} & \textbf{Type}\\\hline
\texttt{p0} & $integer$ &
\texttt{n1} & $integer$\\\hline
\texttt{n2} & $integer$ &
\texttt{X} & $matrix(double)$\\\hline
\texttt{Y} & $matrix(double)$ &
\texttt{b} & $vector(double)$\\\hline
\texttt{nx} & $integer$ &
\texttt{ny} & $integer$\\\hline
\texttt{p} & $integer$ &
\texttt{ctr} & $integer$\\\hline
\texttt{i} & $integer$ &
\texttt{j} & $integer$\\\hline
\texttt{posX} & $integer$ &
\texttt{posY} & $integer$\\\hline
\texttt{total} & $double$ &
\texttt{k} & $integer$\\\hline
\texttt{ans} & $vector(double)$ &
\texttt{dist} & $vector(double)$\\\hline
\end{tabular}}
\label{tab:types-euclidean}
\end{table}

\FloatBarrier

\section{Program Listings}

Here we list the {\R} codes for the remaining programs that we considered for our evaluation. 

\begin{lstlisting}[frame=single,frameround=tttt,caption={Binary Search~\protect\cite{matloff:art-programming:book:2011}},label=lst:binsrch]
binsearch <- function(x,y) {
  n <- length(x)
  lo <- 1
  hi <- n
  while(lo+1 < hi) {
    mid <- floor((lo+hi)/2)
    if (y == x[mid]) return(mid)
    if (y < x[mid]) hi <- mid else lo <- mid
  }
  if (y <= x[lo]) return(lo)
  if (y < x[hi]) return(hi)
  return(hi+1)
}

nn=1e6
x <- sort(sample (1:nn,nn,replace=TRUE))
y <- sample (1:nn,nn,replace=TRUE)
z <- numeric(nn)
system.time(for(i in 1:length(y)) z[i] <- binsearch(x,y[i]))
\end{lstlisting}

\clearpage
\begin{lstlisting}[frame=single,frameround=tttt,caption={Exponential Smoothing~\protect\cite{wickham:advanced-r:book:2014}},label=lst:exps]
exps <- function(x,alpha) {
  s <- numeric(length(x) + 1)
  for (i in seq_along(s)) {
    if (i==1) {
      s[i] <- x[i]
    } else {
      s[i] <- alpha * x[i-1] + (1-alpha) * s[i-1]
    }
  }
  s
}

n <- 1e7
x <- runif(n)
system.time(exps(x,0.5))
\end{lstlisting}

\begin{lstlisting}[frame=single,frameround=tttt,caption={OddCount~\protect\cite{matloff:art-programming:book:2011}},label=lst:oddcnt]
oddcount <- function(x) {
  k <- 0L
  for (n in x) {
    if (n %% 2 == 1) k <- k+1
  }
  return(k)
}

n <- 1e8
x <- sample(1:1000,n,replace=TRUE)
system.time(b<-oddcount(x))
\end{lstlisting}

\begin{lstlisting}[frame=single,frameround=tttt,caption={Discrete Value Time Series, version A~\protect\cite{matloff:art-programming:book:2011}},label=lst:dvts_a]
preda <- function(x,k) {
  n <- length(x)
  k2 <- k/2
  pred <- vector(length=n-k)
  for(i in 1:(n-k)) {
    if(sum(x[i:(i+(k-1))]) >= k2) pred[i] <- 1 else pred[i] <- 0
  }
  return(mean(abs(pred-x[(k+1):n])))
}

n <- 1e7
y <- sample(0:1,n,replace=T)
system.time(preda(y,1000))
\end{lstlisting}\vspace{-2pt}

\newpage
\begin{lstlisting}[frame=single,frameround=tttt,caption={Discrete Value Time Series, version B~\protect\cite{matloff:art-programming:book:2011}},label=lst:dvts_b]
predb <- function(x,k) {
  n <- length(x)
  k2 <- k/2
  pred <- vector(length=n-k)
  sm <- sum(x[1:k])
  if(sm >= k2) pred[1] <- 1 else pred[1] <- 0
  if(n-k >= 2) {
    for(i in 2:(n-k)) {
	  sm <- sm + x[i+k-1] - x[i-1]
	  if(sm >= k2) pred[i] <- 1 else pred[i] <- 0
	}
  }
  return(mean(abs(pred-x[(k+1):n])))
}

n <- 1e7
y <- sample(0:1,n,replace=T)
system.time(predb(y,1000))
\end{lstlisting}\vspace{-2pt}

\begin{lstlisting}[frame=single,frameround=tttt,caption={Euclidean Distance~\protect\cite{templelang:advanced-compilation-tools:ss:2014}},label=lst:euclidean]
dist=function(X, Y) {
  nx = nrow(X)
  ny = nrow(Y)
  p = ncol(X)
  ctr = 1L
  ans = numeric(nx * ny)
  for(i in 1:nx) {
    for(j in 1:ny) {
      posX = i
      posY = j
      total = 0.0
      for(k in 1:p) {
        total = total + (X[posX] - Y[posY])^2
        posX = posX + nx
        posY = posY + ny
      }
      ans[ctr] = sqrt(total)
      ctr = ctr + 1L
    }
  }
  return(ans)
}

p0 = 40L
n1 = 8000L
n2 = 1000L
X = matrix(rnorm(n1 * p0), n1, p0)
Y = matrix(rnorm(n2 * p0), n2, p0)
system.time(b <- dist(X, Y))
\end{lstlisting}\vspace{-2pt}

\end{document}